\documentclass[authoryear,preprint,review,12pt]{elsarticle}

\usepackage{amsmath,amssymb}
\usepackage{graphicx,psfrag,epsf}
\usepackage{enumerate}
\usepackage{natbib}
\usepackage{algorithm}
\usepackage{algpseudocode}
\usepackage{multirow}
\usepackage{url} 
\usepackage[colorlinks,
linkcolor=red,
anchorcolor=blue,
citecolor=green
]{hyperref}
\usepackage[font=normalsize]{caption}
\usepackage{makecell}
\usepackage{rotating}
\usepackage{subcaption}

\def\spacingset#1{\renewcommand{\baselinestretch}%
{#1}\small\normalsize} \spacingset{1}

\usepackage{lineno}

\journal{arXiv}

\begin{document}

\begin{frontmatter}



\title{Dynamic CVaR Portfolio Construction with Attention-Powered Generative Factor Learning}


\author[label1]{Chuting Sun}
\ead{tingsct70@ruc.edu.cn}
\author[label2]{Qi Wu}
\ead{qiwu55@cityu.edu.hk}
\author[label1]{Xing Yan\corref{cor1}}
\cortext[cor1]{Corresponding author.}
\ead{xingyan@ruc.edu.cn}
\address[label1]{Institute of Statistics and Big Data, Renmin University of China,
            Beijing,
            China}
\address[label2]{School of Data Science, City University of Hong Kong,
            Hong Kong SAR}

\begin{abstract}
The dynamic portfolio construction problem requires dynamic modeling of the joint distribution of multivariate stock returns. To achieve this, we propose a dynamic generative factor model which uses random variable transformation as an implicit way of distribution modeling and relies on the Attention-GRU network for dynamic learning and forecasting. The proposed model captures the dynamic dependence among multivariate stock returns, especially focusing on the tail-side properties. We also propose a two-step iterative algorithm to train the model and then predict the time-varying model parameters, including the time-invariant tail parameters. At each investment date, we can easily simulate new samples from the learned generative model, and we further perform CVaR portfolio optimization with the simulated samples to form a dynamic portfolio strategy. The numerical experiment on stock data shows that our model leads to wiser investments that promise higher reward-risk ratios and present lower tail risks.

\end{abstract}




\end{frontmatter}


\section{Introduction}
\label{sec1}

In the portfolio construction problem, one seeks optimal weights invested in stocks and aims for a favored portfolio with a high reward and a low risk. A well-known model-free portfolio strategy is the equal weight (EW) diversification method which requires no extra work to allocate the weights. Actually, \citet{demiguel2009optimal} claimed that there was no method consistently outperforming this naive diversification method on all datasets. Besides, the passive index-tracking method also serves as an effective strategy and as a benchmark \citep{gaivoronski2005optimal,beasley2003evolutionary,sant2017index}. Other methods for dynamic investing, however, rely on dynamic forecasting of the future stock returns or require time-dependent model specifications (if a model is used to describe the stock returns). There have been enduring efforts to model the dynamic evolution of stock returns, most of which rely on some distributional assumptions.

For example, with the multivariate Gaussian distribution assumption, some time series models are able to capture the dynamic patterns of returns, especially the time-dependent structure of the covariance matrix. The most popular model for dynamic covariance matrix might be the dynamic conditional correlation (DCC) model proposed by \citet{engle2002dynamic}. DCC model fits a GARCH(1,1) process to the individual volatility and assumes a similar auto-regressive evolution for the conditional correlation matrix. Despite the clever formulation of DCC model, it still suffers from the curse of dimensionality due to the estimation error of the high-dimensional covariance matrix. Several shrinkage regularization strategies have been proposed to relieve the estimation error, such as the linear \citep{ledoit2004honey}  and non-linear \citep{engle2019large} shrinkage methods that decompose the covariance matrix and control the sparsity of the parameter space. Especially, \citet{engle2012dynamic} simply assumed an equal correlation for all pairs of stocks. 


The introduction of factor structure lightens the worry of the estimation error in moderate-to-high dimensional portfolio constructions. With several common factors, the covariance matrix is reduced to a low-rank representation which provides robustness against the model uncertainty in generalization. Most importantly, the factor structure is handy to combine with more specific models, which further improves the estimation performance. For example, \citet{de2021factor} applied the DCC model to the residuals of an approximate factor model (AFM) and discussed the effect of the number of factors, which turns out that a single factor model is sufficient to achieve satisfactory performance. Moreover, \citet{glasserman2013robust} analyzed the robustness against the uncertainty of a dynamic portfolio control problem based on stochastic factor constructions. Furthermore, \citet{corielli2006factor} proposed a factor-based index tracking method and \citet{creal2015high} incorporated the factor structure with a flexible copula model in capturing the dynamic dependence among assets.


Another branch of literature focuses on constructing black-box neural networks in various topics in finance.
Actually, the popularity of neural networks and deep learning has already brought new perspectives to stock return forecasting or risk forecasting. For example, the Long Short-Term Memory (LSTM) network could learn the long-range dependence from sequential returns, and it is flexible to combine with other methods for better interpretability, as in \citet{fischer2018deep} and \citet{wu2019capturing}. For instance, \citet{wu2019capturing} adopted LSTM to learn the quantile dynamics of financial return series and capture the tail risk.
In the area of financial derivatives, \citet{nian2021learning} proposed a robust Gated Recurrent Unit model with encoder and decoder layers for option hedging. Other works using LSTM or GRU for option hedging include \citet{zhang2021option,carbonneau2021deep,dai2022evaluation}.
Besides, deep learning was applied to portfolio optimization as an end-to-end black-box tool, such as in \cite{zhang2020deep,zhang2021universal}.
There were very few works that adopted deep learning to forecast the core elements of portfolio optimization, such as the covariance matrix \citep{ni2021forecasting}. Most research works were based on traditional linear econometrics models for dynamic covariance forecasting and portfolio optimization \citep{giamouridis2007hedge,de2021factor}. However, \cite{ni2021forecasting} has concluded that deep learning outperforms traditional linear econometrics models.

This paper combines the two branches of research works and proposes a new dynamic factor model for dynamic CVaR portfolio construction. We extend the generative factor model recently proposed in \citet{yan2019cross} (from the machine learning community) to a dynamic version, by combining it with an Attention-GRU network for sequential learning. Such an extension greatly increases the difficulty of the modeling because the conditional joint distribution of stock returns needs to be captured now instead of the unconditional one. The technical challenge is overcome by us with many novel designs.
We accept the common point of view that a single factor in modeling financial returns is enough for achieving good portfolio performance. 
Most crucially, our generative single-factor model incorporates the asymmetric heavy-tail properties of stock returns. Besides, we also take the advantage of the long-term memory capacity of modern neural networks to improve the performance of sequential prediction. Specifically, we combine the Gated Recurrent Unit (GRU) network proposed by \citet{cho2014properties} with the Bahdanau attention mechanism illustrated in \citet{bahdanau2015neural}. The resulting GF-AGRU model named by us is able to capture the complicated dynamic dependence among multivariate stock returns. To tackle the challenge of training such a model, we propose a two-step iterative algorithm to minimize the specified loss functions. After training, the time-varying model parameters, or the time-varying/conditional joint distribution of returns, can be forecasted.

The classical CVaR portfolio optimization formulation in practice (called SAA, see Section \ref{preliminary} for an introduction) needs some samples representing the joint distribution of returns. 
Existing works used the realized returns observed directly in the market.
Our dynamic generative factor model GF-AGRU is basically different. We easily simulate new samples from the predicted conditional joint distribution of monthly returns and then optimize the CVaR portfolio objective with the simulated samples. Then the corresponding optimal weights are assigned to the out-of-sample stock returns in the next month, thus a dynamic portfolio strategy can be constructed. We measure the rewards and various risks of the portfolios given by different methods, under different target return constraints and different confidence levels of CVaR. 
The numerical results of the portfolio performance on the component stocks of the Dow Jones Industrial Average index  exhibit the consistent outperformance of our GF-AGRU model. 
Besides, the superior performance shows adequate robustness under various settings and kinds of randomness. Through analysis, we find that the attention mechanism and the heavy-tail properties are the two key features making our approach successful.

The rest of this paper is organized as follows. We state some preliminaries in Section \ref{preliminary}. Section \ref{methodology} introduces the mathematical formulation of the GF-AGRU model and the comprehensive two-step training algorithm of it. We briefly describe some other competing models in Section \ref{competing-methods}. The details of the numerical experiment and the analysis for the results are covered in Section \ref{numerical experiment}. Finally, we conclude the whole paper in Section \ref{conclusion}.

\section{Preliminaries}\label{preliminary}

\subsection{Static Allocation and Dynamic Allocation}

Assuming that the investor aims to decide the optimal holding weights $w_t\in\mathbb{R}^N$ on $N$ assets satisfying $w_t\ge 0$ and $\mathbf{1}_N^\top w_t = 1$, we let $Y^t\in\mathbb{R}^N$ denote the random vector of asset returns at time $t$, then the portfolio return is $R_t=w_t^\top Y^t$. The static allocation treats the multivariate distribution of $Y^t$ as unchanged in a given period, then estimates the empirical distribution from limited observations of historical returns.  On the contrary, the dynamic portfolio construction assumes that the multivariate/joint distribution of returns is time-varying, and thus one should spare no effort to predict  the conditional joint distribution of returns at every necessary $t$ given past information. We know that the financial markets exhibit dynamic patterns, such as the dynamic volatility which shows clustering or self-exciting. Accordingly, we consider the dynamic allocation problem in this work and update the optimal holding weights $w_t$ at each investment date $t$. So, the forecasting of the distribution of $Y^t$ given past return information will be the key problem to be addressed.

\subsection{The CVaR Portfolio Optimization} 
\label{sec:cvar}

The constructed portfolio is desired to possess a relatively high reward with a low risk, especially that the risk-averse investors tend to place more emphasis on controlling the risk. The classic portfolio optimization considers a mean-variance \citep{markowitz} optimization framework that searches for a trade-off between the reward and the variance risk. However, the mean-variance optimization is more suitable for normally distributed returns, and it fails to find the proper portfolios for heavy-tailed returns. We thus consider the Conditional Value-at-Risk (CVaR) as the risk measure which concerns the down-tail loss of a portfolio. CVaR is a coherent risk measure commonly used in risk management and portfolio optimization, as in \citet{zhu2009worst} and \citet{ban2018machine}. Actually, the $q$-level CVaR of the weighted portfolio return $R_t=w_t^\top Y^t$ is defined as 
\begin{equation}\label{cvar_init}
    \text{CVaR}_q(-w_t^\top Y^t)=\mathbb{E} \left[-w_t^\top Y^t|-w_t^\top Y^t\geq \text{VaR}_q(-w_t^\top Y^t)\right],
\end{equation}
in which $\text{VaR}_q(-w_t^\top Y^t)$ is the $q$-quantile (e.g., 95\%-quantile) of $-w_t^\top Y^t$, or the so-called Value-at-Risk (VaR). For simplicity, we drop the subscript $t$ here. As in \citet{rockafellar2000optimization}, CVaR can be computed as a minimization problem:
\begin{equation}
    \text{CVaR}_q(-w^\top Y) = \min_{\alpha}\left\{\alpha + \frac{1}{1-q}\mathbb{E}\left[(-w^\top Y-\alpha)^+\right]\right\}.
\end{equation}

If we have known the distribution of the random vector $Y$, we can use CVaR as the objective to perform the optimization for the weights $w$ and construct the portfolio:
\begin{equation}
\begin{aligned}
    \min_{w\in\mathbb{R}^N}~ & \text{CVaR}_q(-w^\top Y) \\
    \text{s.t. } & \mathbf{1}_N^\top w = 1 \\
         & w^\top \mu = R_0 \\
         & w\ge 0,
\end{aligned}
\end{equation}
where $\mu=\mathbb{E}[Y]$ and $R_0$ is the target return of the portfolio specified by us. In the above problem, generally we do not know the exact distribution of $Y$. Instead we have samples of $Y$: $\tilde{Y}^1,\dots,\tilde{Y}^n$, no matter whether they are historically observed or newly simulated from a distribution. In this case, the sample approximate average (SAA, see \citet{ban2018machine}) approach to this problem is
\begin{equation}\label{eqn:CVaR-SAA}
\begin{aligned}
    \min_{w\in\mathbb{R}^N}~ & \widehat{\text{CVaR}}_q(-w^\top Y) \\
    \text{s.t. } & \mathbf{1}_N^\top w = 1 \\
         & w^\top \hat{\mu} = R_0 \\
         & w\ge 0,
\end{aligned}
\end{equation}
where $\hat{\mu}=\frac{1}{n}\sum_{j=1}^n \tilde{Y}^j$ and
\begin{equation}\label{eqn:CVaR-empirical}
    \widehat{\text{CVaR}}_q(-w^\top Y) = \min_{\alpha}\left\{\alpha + \frac{1}{(1-q)n}\sum_{j=1}^n(-w^\top \tilde{Y}^j-\alpha)^+\right\},
\end{equation}
which is a sample average estimator for $\text{CVaR}_q(-w^\top Y)$. However, in dynamic setting, we usually only have one single observation of multivariate stock returns at each time $t$. If we can have a good forecast of the distribution of $Y^t$ based on past information, we can simulate many new samples of $Y^t$ and solve the above SAA formulation to obtain an optimal portfolio. This is the main purpose of this paper, compared to the static allocation case where historical observations of returns are used in Equation \eqref{eqn:CVaR-SAA} (\textit{i.i.d.} assumption, equivalently).


\subsection{Generative Machine Learning Models}

In machine learning, the terminology \textit{generative models} refer to those models which can be used to simulate new samples, opposed to \textit{discriminative models}. Generative models are an important branch of machine learning or deep learning, for example, the well-known Generative Adversarial Networks (GANs, see \citet{goodfellow2020generative}) and Wasserstein-GAN \citep{arjovsky2017wasserstein} belong to this branch. Most recent attractive generative models have the following form:
\begin{equation}\label{eqn:gml.models}
Y=G(Z;\Theta),
\end{equation}
where $Y$ is the target random variable (vector) with samples observed, such as multivariate stock returns. $Z$ is called a base variable (vector) whose distribution is usually simple, such as Gaussian or uniform. $G$ is a transformation that is used to convert the simple distribution to the complicated distribution (of $Y$) we are interested in. 
$\Theta$ is the parameters of $G$ that need to be learned or optimized.
The aim is to approach the distribution of $Y$ with finite samples observed.
One can see, it is quite easy to simulate new samples of $Y$ through \eqref{eqn:gml.models}, compared to the traditional density function approach for describing a distribution. One only needs to generate samples of $Z$ and then apply $G$ on them.

Examples of popular generative models taking the form \eqref{eqn:gml.models} include variational auto-encoders (VAEs, see \citet{kingma2019introduction}), various GANs, and Normalizing Flows \citep{kobyzev2020normalizing,papamakarios2021normalizing}.
They are proposed or designed with different ideas and considerations. But the purposes are the same, to describe or approach the distribution of $Y$ and to easily sample.
Another commonality is that all of them adopt a neural network for $G$, making the approaches nonparametric.
VAEs and GANs have shown their successes mainly on image generating and editing. Recently, \citet{cont2022tail} proposed Tail-GAN for nonparametric scenario generation with tail risk estimation. Wasserstein-GAN was applied for causal inference in \citet{athey2021using}, used for simulating datasets with counterfactuals and evaluating causal inference methods.

Normalizing Flows are the models that require $Y$ and $Z$ have a same dimension and $G$ has an inverse in \eqref{eqn:gml.models}. So, the density of $Y$ can be computed by the change-of-variable technique. 
Conveniently, Normalizing Flow models can be learned with maximum likelihood, while GANs must be trained with adversarial schemes. 
In this paper, our generative model is a specific Normalizing Flow model tailored for financial return modeling. It does not adopt a neural network for $G$. Instead, a carefully designed $G$ with financial market considerations is adopted. Most importantly, it is dynamic with the assistance of an Attention-GRU network, which is a powerful sequential learning model. Originating from the recent machine learning community, an earlier version \citep{yan2019cross} of it has been proven to be successful.

\subsection{The GRU Network and Attention Mechanism}
Deep learning has been successfully applied in finance, such as in empirical asset pricing or cross-section of stock returns \citep{gu2020empirical,chen2023deep}.
In deep learning, recurrent neural networks (RNNs) are extremely useful in dealing with sequential prediction problems. Especially, the Long Short-Term Memory (LSTM, see \citet{hochreiter1997long}) and its younger sibling Gated Recurrent Unit (GRU, see \citet{cho2014properties}) are known for their powerful capability in capturing the time-dependent structure and long memory in the sequential data. \citet{zhai2020neural} compared the performance of the LSTM model and the GRU model in volatility prediction. The results show that these two models lead to comparable accuracies, while GRU behaves more smoothly in convergence. In this work, we adopt the GRU network in our model construction. 

The attention mechanism further helps hold necessary information adaptively in dealing with long sequences, whose merits have been established in machine translation \citep{vaswani2017attention} and financial forecasting \citep{chen2019exploring}. Especially in machine translation, the attention mechanism has achieved great successes. And it is also one of the foundations of ChatGPT \citep{openai2023gpt4} and its predecessors. 
We utilize a location-based attention structure proposed by \citet{bahdanau2015neural} which reweights the outputs of the recurrent neural network with an alignment model. Let $T$ denote the length of the input sequence and $y_{1:T}$ be the outputs of a GRU, then we simply compute a new output $\tilde{s} =\sum_{\tau=1}^{T} \alpha_{\tau} y_{\tau}$ with some soft weights $\alpha_\tau$, $\tau=1,\dots,T$. Combined with the single-layer GRU network, the weights $\alpha_{1:T}$ are calculated as
\begin{equation}
    \begin{aligned} 
    \alpha_{1:T} &=\text{softmax}(e_{1:T}),\\
    e_\tau &=v_{a}^{\top} \tanh \left(W_{a} y_{\tau}+U_{a} y_{T}\right),
    \end{aligned}
\end{equation}
where $y_{T}$ is also the final vector of hidden states. Denoting $n$ as the number of hidden units in GRU, then $v_a\in \mathbb{R}^{n}$ and $W_a,U_a \in \mathbb{R}^{n \times n}$ are weight matrices that act as model parameters of the attention layer. After that, a linear layer is then applied on the new output $\tilde{s}$. We call this whole architecture the Attention-GRU network (AGRU).

\section{Methodology}
\label{methodology}

\subsection{The Naive Factor Model}\label{sec-naive-factor}

Before describing our proposed method in detail, we first review the construction of the naive factor model.
Let $r_t$ denote an $N$-dimensional vector of stock returns and $f_t$ be the $K$-dimensional common factors, then a $K$-factor model is formulated as
    $r_t=\alpha_t+B_t f_t+\epsilon_t$, 
where $\alpha_t$ is the vector of intercepts and $B_t$ is an $N\times K$ matrix that contains corresponding factor loadings. The unexplained residual vector $\epsilon_t$ is usually assumed to be normally distributed with zero mean and covariance matrix $\Omega_t$. We refer to a model with a diagonal $\Omega_t$ as an exact factor model (EFM), and otherwise it is called an approximate factor model (AFM). 

In this work we assume the stock returns conform to the EFM structure and only consider the case of the single factor model with $K=1$, as suggested by \citet{de2021factor}.
We denote $Y_M^t$ as the future one-month market return and $Y_i^t$ as the future one-month return of the $i^{\text{th}}$ individual stock, $i=1,\dots, N$, at an investment date $t$. Then the naive factor model can be expressed in such a specific form:
\begin{equation}\label{naive-factor}\tag{naive factor}
\begin{aligned}
    Y_M^t &= \alpha_M^t + \beta_M^t Z_M^t,\\
    Y_i^t &= \alpha_i^t + \beta_i^t Z_M^t+\gamma_i^t Z_i^t,
\end{aligned}
\end{equation}
where $Z_M^t$ is the latent market factor, $Z_i^t$ is the unexplained residual of the $i^{\text{th}}$ stock, assuming that $ Z_M^t$ and $Z_i^t$ follow the standard normal distribution independently. 
Noticing that the Naive Factor model can also be considered as a simplified multivariate Gaussian distribution.
$\alpha_M^t$ is the expectation of market return, and $\alpha_i^t$ is the expectation of the $i$-th stock return.
$\beta_M^t$ is the volatility of the market return, $\beta_i^t$ is the coefficient to the market factor, and $\gamma_i^t$ is the volatility of the unexplained residual. We allow the coefficient $\beta_i^t$ to vary with time, which is different from the constant-coefficient assumption in most existing works, e.g., \citet{engle2012dynamic} and \citet{levy2021dynamic}. Finally, we hope to use an Attention-GRU network to predict these time-varying parameters $\Theta^t_{\text{naive}}= \{\alpha_M^t,\alpha_i^t,\beta_M^t,\beta_i^t,\gamma_i^t\}$:
\begin{equation}
    \Theta^t_{\text{naive}} = \text{AGRU}(F_{<t};\theta_{\text{AGRU}}),
\end{equation}
where $F_{<t}$ is the features extracted from past observations containing predictive information, and $\theta_{\text{AGRU}}$ is the parameters of AGRU that need to be optimized. We show in the next section that the naive factor model with Attention-GRU (abbreviated as Naive-AGRU) is a special case of our proposed model.

\subsection{The Dynamic Generative Factor Model with Tail Properties}

We extend the non-linear transformation proposed in \citet{yan2019cross} to a dynamic version and extend the naive factor model to incorporate a better depiction of tail properties of stock returns. The proposed dynamic generative factor model is constructed as 
\begin{equation} \label{gx-factor}\tag{generative factor}
\begin{aligned}
Y_M^t &= \alpha_M^t +\beta_M^t g(Z_M^t;u_M,v_M),\\
Y_i^t &= \alpha_i^t +\beta_i^t g(Z_M^t;u^M_i,v^M_i)+\gamma_i^t g(Z_i^t;u_i,v_i),
\end{aligned}
\end{equation}
where  $Z_M^t$ and $Z_i^t$ still follow standard normal distribution independently. The $g(\cdot)$ function in the \ref{gx-factor} model is specified as
\begin{equation}\label{gx-func}
    g(x;u,v) = x(\frac{u^x+v^{-x}}{A}+1),
\end{equation}
in which $A$ is a positive scaling constant and we set $A=4$ as in \citet{yan2019cross}. 
The strictly monotonically increasing $g(\cdot)$ function\footnote{Please see \cite{wu2019capturing} for the proof, and we can easily derive that $g^{-1}$ uniquely exists and thus the transformation defined by our \ref{gx-factor} model has an inverse.} transforms $Z_M^t$ or $Z_i^t$ to a more heavy-tailed random variable whose right and left tails are decided by the parameters $u,v\ge 1$.
We refer $\nu=\{u,v\}$ to time-invariant tail parameters. More specifically, $\nu_M=\{u_M,v_M\}$, $\nu_i^M=\{u_i^M,v_i^M\}$, and $\nu_i=\{u_i,v_i\}$. 
Larger tail parameters lead to relatively heavier tails. Particularly, when all $u=v=1$, the generative factor model reduces to a naive factor model. In \ref{gx-factor}, $\nu_M$, $\nu_i^M$, and $\nu_i$ can all be different, leading to much flexible tail properties for stock returns.

We further use an Attention-GRU network to forecast the time-varying parameters $\Theta^t_{\text{GF}}= \{\alpha_M^t,\alpha_i^t,\beta_M^t,\beta_i^t,\gamma_i^t\}$:
\begin{equation}
    \Theta^t_{\text{GF}} = \text{AGRU}(F_{<t};\theta_{\text{AGRU}}),
\end{equation}
where $F_{<t}$ is the features extracted from past observations containing predictive information, and $\theta_{\text{AGRU}}$ is the parameters of AGRU that need to be optimized. While $Y_M^t$ and $Y_i^t$ are future one-month returns, $F_{<t}$ can be constructed from historical daily returns to incorporate as much as information. Besides, the training set collects data points on a daily frequency, so adjacent $Y_M^t$ (or $Y_i^t$) will have overlaps in periods, which can increase the training set size for better learning. 
During the training, the time-invariant tail parameters $\{\nu_M,\nu_i^M,\nu_i\}$ will be learned or optimized too. We abbreviate the whole model as GF-AGRU.

\subsection{Loss Functions}

Machine learning models need loss functions as learning objectives or optimization objectives.
Given the \ref{gx-factor} model and $S$ samples of monthly returns $Y_M^t,~Y_i^t$, $i=1,\dots,N$, $t=1,\dots,S$, we have the likelihood 
\begin{equation}
\begin{aligned}
p(Y_M^t, Y_1^t, \dots, Y_N^t) & = p(Y_M^t)p(Y_1^t, \dots, Y_N^t|Y_M^t) \\
& = p(Y_M^t)p(Y_1^t|Y_M^t)\cdots p(Y_N^t|Y_M^t),
\end{aligned}
\end{equation}
because of the conditional independence (given $Y_M^t$, then $Z_M^t$ is determined). Now we are able to further deduce the form of the negative log-likelihood (NLL) loss as follows:
\begin{equation}\label{losssum}
\text{NLL} = \sum_{t=1}^S -\ell^t =\sum_{t=1}^S(-\ell_M^t)+ \sum_{t=1}^S \sum_{i=1}^N (-\ell_{i|M}^t),
\end{equation}
where $\ell_M^t$ is the log-likelihood of market return at time $t$, and $\ell_{i|M}^t $ is the quasi-log-likelihood of the $i^{\text{th}}$ stock return conditional on the realizations of market factor. Because of the formulation in \ref{gx-factor} and the change-of-variable technique, we can have
\begin{align}
    \ell_M^t & = \log p(Y_M^t) = -\log (\beta_M^t g'(\tilde{Z}_M^t;\nu_M))+\log \phi(\tilde{Z}_M^t), \label{MLL}\\
    \ell_{i|M}^t & = \log p(Y_i^t|Y_M^t)=  -\log (\gamma_i^t g'(\tilde{Z}_i^t;\nu_i))+\log \phi(\tilde{Z}_i^t),\label{quasi-LL}
\end{align}
in which $\phi$ is the density of standard Gaussian distribution and $\tilde{Z}_M^t, \tilde{Z}_i^t$ are the realized latent market factor and the realized residual term at time $t$:
\begin{align}
       \tilde{Z}_M^t &= g^{-1}(\frac{Y_M^t-\alpha_M^t}{\beta_M^t};\nu_M),\label{latentfac}\\
       \tilde{Z}_i^t &= g^{-1}(\frac{Y_i^t-\alpha_i^t-\beta_i^tg(\tilde{Z}_M^t;\nu_i^M)}{\gamma_i^t};\nu_i).\label{latentstock}
\end{align}
Note that $g'(x;\nu)$ and $g^{-1}(\cdot;\nu)$ are respectively the derivative with respect to $x$ and the inverse of the $g(\cdot)$ function in Equation (\ref{gx-func}). 

We observe that the minimization of the NLL loss can be decomposed into two steps: 
\begin{enumerate}
\item Firstly minimize $\sum_{t=1}^S (-\ell_M^t)$ with respect to $\nu_M$ and $\Theta_M^t=\{\alpha_M^t,\beta_M^t\}$ and calculate the realized latent market factor $\tilde{Z}_M^t$, given known $Y_M^t$ for all $t$;
\item Then fix $\tilde{Z}_M^t$ for all $t$ and minimize the quasi-NLL $\sum_{t=1}^S\sum_{i=1}^N (-\ell_{i|M}^t)$ with respect to $\{\nu_i^M,\nu_i\}$ and $\Theta_i^t = \{\alpha_i^t,\beta_i^t,\gamma_i^t\}$ for all individual stocks. 
\end{enumerate}
We thus split the whole model training into these two steps, and to reduce the difficulty of training, we separately apply a Attention-GRU network to learn and predict ($\Theta_M^t$ or $\Theta_i^t$) in each of the two steps.
We assume all tail parameters $\nu_M,\nu^M_i,\nu_i$ are fixed over the whole period but allow time-varying $\Theta^t_M,\Theta^t_i$ to incorporate the temporal dynamics of returns. Except for optimizing the AGRU parameters, to learn the time-invariant $\nu_M$ as well, we design an iterative algorithm in Step 1 to learn $\nu_M$ and $\Theta^t_M$ alternately, as introduced in the following subsection. A similar algorithm is also designed for learning $\{\nu^M_i,\nu_i\}$ and $\Theta^t_i$ alternately in Step 2. 

\subsection{The Two-Step Training with Alternately Updating} \label{twostep}

We name the two steps of the training as market return fitting and individual stock return fitting. In each step, the time-invariant parameters $\nu$ and the neural network parameters of Attention-GRU (used to predict time-varying $\Theta^t_M$ or $\Theta^t_i$) are updated iteratively and alternately. We summarize the two steps of the training in Algorithm \ref{algorithm_N} and Algorithm \ref{algorithm_i} respectively.

\subsubsection{Market Return Fitting}

In this step, we construct an algorithm to minimize the negative log-likelihood $\sum_{t=1}^S (-\ell_M^t)$ and learn $\nu_M$ and $\Theta^t_M$, as described in Algorithm \ref{algorithm_N}. 
Note that the Attention-GRU network will take in a historical window of \emph{daily} market returns $r_M^{\tau}$, $\tau=1,\dots, T$ as inputs and output time-varying parameters $\Theta_M^t$ for the \emph{monthly} return $Y^t_M$. We also add the squared terms to the inputs. That is, the final feature matrix $F_{<t}$ is
\begin{equation}\label{eqn:features_Ft}
    F_{<t} = \left[
    \begin{array}{ccc}
        r_M^{1} & \cdots & r_M^{T} \\
        (r_M^{1}-\bar{r}_M)^2 & \cdots & (r_M^{T}-\bar{r}_M)^2
    \end{array}
    \right],\quad
    \bar{r}_M = \frac{1}{T}\sum_{\tau=1}^T r_M^{\tau},
\end{equation}
where $T$ is the length of the lookback window and $\bar{r}_M$ is the historical mean of the daily return sequence. Again the training set collects data points on a daily frequency, so adjacent $Y_M^t$ (or $Y_i^t$) will have overlaps in periods, which can increase the training set size for better learning. 

Then we construct two sub-procedures to learn the time-invariant tail parameters $\nu_M$  and the Attention-GRU network parameters $\theta_{\text{AGRU}}$ alternately, named as FIX-OPTIM and TV-AGRU. These two sub-procedures will aim at optimizing the same likelihood loss $\sum_{t=1}^S (-\ell_M^t)$, but differ in specific gradient back-propagation and descent. In each of the sub-procedures, we update one kind of parameters with some steps when keeping the other fixed. We then iteratively conduct these two sub-procedures until reaching the stopping condition for convergence. After that, we calculate the in-sample realized latent market factor $\tilde{Z}_M^t$ for all $t$ using Equation (\ref{latentfac}). 
Specifically, we constrain the output of FIX-OPTIM to ensure $1\leq \nu_M\leq 3$ and set proper bounds for the outputs of TV-AGRU to increase the stability in gradient descent. 
We construct a self-defined auto-gradient unit for the inverse of $g(\cdot)$ function based on the flexible modules provided by PyTorch.  Besides, we apply the \emph{RMSProp} optimizer with momentum $m=0.2$, which is a common choice in deep learning for financial studies, as discussed in \citet{sezer2020financial}. 

\subsubsection{Individual Stock Return Fitting}

Given the calculated in-sample $\tilde{Z}_M^t$ in Step 1, we further construct Algorithm \ref{algorithm_i} to minimize the quasi-NLL $\sum_{t=1}^S \sum_{i=1}^N (-\ell_{i|M}^t)$ and learn the parameters $\{\nu^M_i,\nu_i\}$ and $\Theta^t_i$ for all individual stocks. Since the parameters of each stock are independent, we divide the task into $N$ individual sub-problems and tackle them separately.
The corresponding unknown parameters for stock $i$ are the time-invariant tail parameters $\{\nu^M_i,\nu_i\}$ and the Attention-GRU network parameters (used to predict $\Theta_i^t=\{\alpha_i^t,\beta_i^t,\gamma_i^t\}$), and the loss function is $\sum_{t=1}^S (-\ell_{i|M}^t)$. Note that for each stock $i$, a separate Attention-GRU will be used and trained, resulting in $N$ Attention-GRU networks totally. Each of them is with a very low dimension, thus the computational cost is not high.

 Similar as Algorithm \ref{algorithm_N} in Step 1, 
here we adopt two sub-procedures FIX-OPTIM and TV-AGRU iteratively too, until reaching a certain stopping criteria. 
The specifications of the two sub-procedures are similar to those in Step 1. The inputs or features $F_{<t}$ of Attention-GRU are similar too, but adding daily returns of the individual stock and the squared terms as well. This results in a input dimension of 4, compared to the dimension 2 in Step 1. 
In both steps, we adopt the widely-used early-stopping strategy in deep learning to stop the training, with a validation set extracted from the training set.
Now the training of the whole GF-AGRU model is finished. As discussed above, the Naive-AGRU model is a special case of the GF-AGRU model when the tail parameters are all one. Therefore, we handily modify Algorithm \ref{algorithm_N} and Algorithm \ref{algorithm_i} by fixing $\nu_M=\nu^M_i=\nu_i=1$, $i=1,\dots, N$ in TV-AGRU and omit the FIX-OPTIM sub-procedure when training a Naive-AGRU model, while other parts are the same as the GF-AGRU model.

\begin{algorithm}
\caption{The market return fitting step in training GF-AGRU.} \label{algorithm_N}
\textbf{Hyperparameters}: learning rate $l_{\text{fix}}$ of the FIX-OPTIM sub-procedure, learning rate $l_{\text{tv}}$ of the TV-AGRU sub-procedure, number of maximum iterative steps $N_m$, number of training epochs $N_{\text{fix}}$ of FIX-OPTIM, number of training epochs $N_{\text{tv}}$ of TV-AGRU, and the training set size $S$.\\
\textbf{Input}: training data including the features $F_{<t}$ and the label $Y_M^t$. $F_{<t}$ is constructed by Equation \eqref{eqn:features_Ft} using historical daily market returns, and $Y_M^t$ is the future one-month return. The training set collects data points on a daily frequency. \\
\textbf{Initialize}: tail parameters $\nu_M=\{u_M,v_M\}$ and the Attention-GRU network parameters $\theta_{\text{AGRU}}$.
\begin{algorithmic}[1]
\For{$b=1:N_m$}
\State (TV-AGRU)
\State Fix $\nu_M$ given by FIX-OPTIM.
\For{$j=1:N_{\text{tv}}$}
    \State Compute $\Theta^t_M = \{\alpha^t_M,\beta^t_M\} = \text{AGRU}(F_{<t};\theta_{\text{AGRU}})$.
    \State Compute the NLL loss $L=\sum_{t=1}^S (-\ell_M^t)$ and its partial derivatives with respect to $\theta_{\text{AGRU}}$.
    \State Update: $\theta_{\text{AGRU}} \gets \text{RMSProp}(\theta_{\text{AGRU}},\nabla_{\theta_{\text{AGRU}}} L,l_{\text{tv}})$.
\EndFor 
\State (FIX-OPTIM)
\State Fix $\theta_{\text{AGRU}}$ given by TV-AGRU.
\State Compute $\Theta^t_M = \{\alpha^t_M,\beta^t_M\} = \text{AGRU}(F_{<t};\theta_{\text{AGRU}})$ and fix $\Theta^t_M$.
\For{$j=1:N_{\text{fix}}$}
    \State Compute the NLL loss $L=\sum_{t=1}^S (-\ell_M^t)$ and its partial derivatives with respect to $\nu_M$.
    \State Update: $\nu_M \gets \text{RMSProp}(\nu_M,\nabla_{\nu_M} L,l_{\text{fix}})$.
\EndFor 
\EndFor
\end{algorithmic}
\textbf{Output}: learnable parameters $\nu_M$ and $\theta_{\text{AGRU}}$; the time-varying model parameters $\Theta^t_M$; the realized latent market factor $\tilde{Z}_M^t$ given by Equation (\ref{latentfac}). 
\end{algorithm}

\begin{algorithm}
    \caption{The individual stock return fitting step in training GF-AGRU (for stock $i$).} \label{algorithm_i}
    \textbf{Hyperparameters}: the same as in Algorithm \ref{algorithm_N}.\\
    \textbf{Input}: training data including the features $F_{<t}$ and the label $Y_i^t$. $F_{<t}$ is constructed by concatenating historical daily market returns and historical daily stock returns (both in the form of Equation \eqref{eqn:features_Ft}), and $Y_i^t$ is the future one-month stock return. The training set collects data points on a daily frequency. \\
    \textbf{Initialize}: tail parameters $\nu^M_i=\{u^M_i,v^M_i\},\nu_i=\{u_i,v_i\}$ and the Attention-GRU network parameters $\theta_{\text{AGRU}}$ (a different Attention-GRU from that in Algorithm \ref{algorithm_N} and those for other stocks).
    \begin{algorithmic}[1]
    \For{$b=1:N_m$}
    \State (TV-AGRU)
    \State Fix $\{\nu^M_i,\nu_i\}$ given by FIX-OPTIM.
    \For{$j=1:N_{\text{tv}}$}
        \State Compute $\Theta^t_i = \{\alpha^t_i,\beta^t_i,\gamma^t_i\} = \text{AGRU}(F_{<t};\theta_{\text{AGRU}})$.
        \State Compute the quasi-NLL loss $L=\sum_{t=1}^S (-\ell_{i|M}^t)$ (with known $\tilde{Z}_M^t$ obtained from Algorithm \ref{algorithm_N}, see Equation \eqref{latentstock}) and its partial derivatives with respect to $\theta_{\text{AGRU}}$.
        \State Update: $\theta_{\text{AGRU}} \gets \text{RMSProp}(\theta_{\text{AGRU}},\nabla_{\theta_{\text{AGRU}}} L,l_{\text{tv}})$.
    \EndFor 
    \State (FIX-OPTIM)
    \State Fix $\theta_{\text{AGRU}}$ given by TV-AGRU.
    \State Compute $\Theta^t_i = \{\alpha^t_i,\beta^t_i,\gamma^t_i\} = \text{AGRU}(F_{<t};\theta_{\text{AGRU}})$ and fix $\Theta^t_i$.
    \For{$j=1:N_{\text{fix}}$}
        \State Compute the quasi-NLL loss $L=\sum_{t=1}^S (-\ell_{i|M}^t)$ (with known $\tilde{Z}_M^t$ obtained from Algorithm \ref{algorithm_N}, see Equation \eqref{latentstock}) and its partial derivatives with respect to $\{\nu^M_i,\nu_i\}$.
        \State Update: $\{\nu^M_i,\nu_i\} \gets \text{RMSProp}(\{\nu^M_i,\nu_i\},\nabla_{\{\nu^M_i,\nu_i\}} L,l_{\text{fix}})$.
    \EndFor 
    \EndFor
    \end{algorithmic}
    \textbf{Output}: learnable parameters $\{\nu^M_i,\nu_i\}$ and $\theta_{\text{AGRU}}$; the time-varying model parameters $\Theta^t_i$.
    \end{algorithm}

\subsection{Computational Complexity}

We indeed decompose a $(N+1)$-dimensional learning problem into $N+1$ separate one-dimensional learning problems. Each asset will have a learning process separately. In each one-dimensional problem, an Attention-GRU network with a low hidden state dimension is enough because the feature dimension of $F_{<t}$ is 2 for the market and is 4 for individual stock. The training of one Attention-GRU will not be costly, especially when we adopt the early stopping strategy. So totally, the computational cost will be satisfactory. 

In TV-AGRU, we denote the computational cost of performing gradient descent one time on one data point as $c_{\text{tv}}$. Similarly, we denote the cost of one gradient descent on one data point in FIX-OPTIM as $c_{\text{fix}}$. Supposing there are totally $N_{\text{tr}}$ data points in training set, then the total cost of training a GF-AGRU is $O\left((N+1)N_m N_{\text{tr}}(N_{\text{tv}}c_{\text{tv}} + N_{\text{fix}}c_{\text{fix}})\right)$. 
For comparison, we define the original problem as minimizing the total loss in Equation \eqref{losssum} directly without the decomposition and with one high-dimensional Attention-GRU network. The inputs and outputs are the aggregations of all inputs and all outputs of our problem, respectively. So, the input dimension will be around 120, because the input dimension in our problem is 4 for each stock. Similarity, the output dimension will be around 90. For the hidden state dimension, we set it to twice the output dimension (similar as in our problem).

When the network size is $N+1$ times the one in our problem, the computational complexity of performing one gradient descent on one data point will be $(N+1)^2$ times (because of the matrix operation). Omitting the $c_{\text{fix}}$ term ($\ll c_{\text{tv}}$), the complexity of the original problem should be $O((N + 1)^2c_{\text{tv}} N_{\text{tr}} N_{\text{tv}})$, which is significantly larger than our $O((N + 1)N_mN_{\text{tr}} N_{\text{tv}}c_{\text{tv}})$. This is because $N_m = 6 <N + 1$ in our experiment. Here, the $(N + 1)N_m$ term is the unique part in our problem, as $N_m$ is the number of iterative steps used to alternate between FIX-OPTIM and TV-AGRU and $N + 1$ is the number of assets. To conclude, our method reduces the computational complexity, compared to the simple and straightforward brute-force method.


\subsection{CVaR Portfolio Construction} \label{cvar_section}


Given the past information up to date $t$, the GF-AGRU model can predict all the parameters of the \ref{gx-factor} model, which describes the joint distribution of the future one-month returns $Y^t_M,Y^t_1,\dots,Y^t_N$ starting at date $t$.
On the new testing set, through this dynamic generative factor model, it is convenient to simulate new independent samples from the predicted factor model. Specifically, at a certain monthly investment date, we simulate $n$ new samples for the future $Y^t_1,\dots,Y^t_N$ and perform the CVaR portfolio optimization with quantile level $q$ introduced in Section \ref{sec:cvar} to find the optimal weights $w_t$.
We then roll forward to next month to predict again and to construct a monthly re-balancing portfolio. Note that the training set collects data points on a daily frequency, while the prediction happens monthly when testing.


Rather than finding the global minimum of the risk frontier, we put on some constraints on the optimization. We assume no short sale for any stock and force these weights of the stocks to be summed as $1$. Moreover, we require that the obtained portfolio should reach a certain target return. Please refer to Equation \eqref{eqn:CVaR-SAA} and Equation \eqref{eqn:CVaR-empirical} for the details. 
The target return $R_0$ is set by us to be around the overall average monthly return of all stocks. In actual implementation, we solve the CVaR portfolio optimization problem with the MATLAB Optimization Toolbox \citep{OptimizationToolbox} and Financial Toolbox \citep{FinancialToolbox} 
to find the optimal weights of the stocks.

\section{Competing Methods}
\label{competing-methods}

We compare the proposed GF-AGRU model with some benchmark models. 
Firstly, we consider two static allocation strategies. One is the naive equal weight (EW) diversification strategy, which assigns time-invariant equal weight to each stock, i.e., $w_{t}=[1/N,\dots,1/N]^\top$. The other is the static SAA of CVaR portfolio optimization, i.e., using historical returns with the \textit{i.i.d.} assumption in Equation \eqref{eqn:CVaR-SAA}.
As to the dynamic competing models, we compare to the DCC model with a factor structure. DCC model is a popular multivariate financial time series model, suitable for the comparison. 

At last, to better understand the function or usefulness of each component of the proposed GF-AGRU, we compare it to two degenerated versions of it. One is the Naive-AGRU model described in Section \ref{sec-naive-factor}, which discards the tail properties of the stock returns in the model, aiming for discovering the importance of tail risk modeling in financial markets. The other is a GF-GRU model in which we drop the attention mechanism/layer in GF-AGRU, aiming for checking the necessity of including attention mechanism in our model.

\subsection{The Factor-DCC model}
It is necessary to briefly introduce the DCC-based model considered. We consider the DCC model with normal distribution assumption. To be consistent with the formulation of GF-AGRU, we adopt a factor-based DCC to reduce the parameter complexity. We still select a market factor as the common factor and assume the correlations among stocks are fully captured by a single factor, thus leading to a diagonal covariance matrix for the independent residuals. \citet{de2021factor} combined an approximate factor structure with a non-linear shrinkage DCC model, but they assume constant factor loadings and directly estimate them with Ordinary Least Square. 

In contrast, we allow the factor loadings to vary with time. We denote the de-meaned market return and de-meaned stock return as $r_M^t$ and $r_i^t$, 
and they have volatilities $\sigma_M^t$ and $\sigma_i^t$ respectively. The Factor-DCC model we adopt assumes that $\sigma_M^t$ and $\sigma_i^t$, $i=1,\dots,N$ all follow a respective GARCH(1,1) process, and meanwhile, the correlation between any stock and the market composes a single-factor structure:
    $\frac{r_i^t}{\sigma_i^t}=\rho^t_{i}\frac{r_M^t}{\sigma_M^t}+\sqrt{1-\left(\rho^t_{i}\right)^2}Z_i^t$,
where $\rho^t_{i} = \mathrm{corr}(r_i^t,r_M^t)$ is the time-varying correlation. The unexplained residuals $Z_i^t$, $i=1,\dots,N$ are assumed to be independent and follow a standard normal distribution.
We assume the correlation $\rho^t_{i}$ also follows a similar evolution process. For simplicity, we denote the $2\times 2$ correlation matrix as $Q^t_i$ and update it with the following formula:
\begin{equation}\label{dcc-rho}
    Q^t_i = \left[
    \begin{array}{cc}
        1 & \rho^t_i \\
        \rho^t_i & 1
    \end{array}
    \right],\quad
    Q^t_i = (1-a-b)\Gamma_i+ a e^{t-1}_i \left(e^{t-1}_i\right)^\top +bQ^{t-1}_i,
\end{equation}
where $e^t_i =[r_M^t/\sigma_M^t,r_i^t/\sigma_i^t]^\top$ and $\Gamma_i$ is a constant symmetric matrix with all-one diagonal elements. 

For the vanilla DCC model, the estimation of the constant matrix $\Gamma_i$ is a major challenge, especially in high-dimensional cases. A traditional estimation method for $\Gamma_i$ is the sample correlation matrix (see \citet{engle2002dynamic}) which is prone to estimation error. \citet{engle2019large} applied a nonlinear shrinkage method to estimate $\Gamma_i$ and obtained well-conditioned estimators in high dimensions. 
However, due to the factor structure of the Factor-DCC model here, we can avoid these troubles because $\Gamma_i$ (a $2\times 2$ matrix) only has one unknown parameter. We only need to assure the diagonal elements of $Q^t_i$ are scaled to ones.

Note that we can train a DCC model with \emph{monthly returns} and directly predict the distribution parameters for the one-step ahead monthly returns, hence predicting the conditional joint distribution. Besides, it is also possible to train the DCC model with \emph{daily returns} and make multi-step ahead simulations, then form a monthly return simulation with 21 consecutive daily simulations. These monthly return simulations can be used as the samples needed in CVaR portfolio optimization. We refer to the DCC model trained with monthly returns as DCC-MM, and the model trained with daily returns as DCC-DM. Our experience indicates that their performance is close on the portfolio construction. So, we will only report the results of DCC-MM in the following experimental section.


\subsection{GANs}

As a prominent paradigm in the domain of generative models, Generative Adversarial Networks (GANs) exhibit potential in financial time series learning and simulation. GANs attempt to generate new data that follows the same distribution as the provided inputs. For instance, GANs are supposed to simulate new multivariate return samples that exhibit the same distribution as the observed returns in the training set. From this perspective, it is crucial to realize that GANs represent a static model rather than a dynamic model, as they are learning the unconditional distribution of the multivariate time series. This contradicts the dynamic objective in this study. Oppositely, at an investment date $t$ (e.g., the end of a month), our model simulates new samples that follow the conditional distribution of future returns, conditional on the information available up to date $t$.
The new samples serve as the inputs to the CVaR optimization problem. To be short and precise, the forecasting of the conditional multivariate distribution is the key to the portfolio optimization problem here.

An exception of GAN-based model is the Time-series GAN or TimeGAN in \cite{yoon2019time}, which modeled the conditional distribution of time series. However, one of the main differences is that the framework of TimeGAN is entirely a black box without careful considerations of dependence structure among asset returns (we use the well-explained factor model) and without careful specifications of tail properties (we use the well-explained $g(\cdot)$ function), both of which are essential for precise distribution forecasting. Besides, TimeGAN is essentially a recurrent neural network which transforms a sequence of pure noises to the target time series. However, as we know, it is unclear how to use it to do conditional distribution forecasting in a new time series.
At last, it adopts a complex embedding from original time series to the latent space, which is redundant for financial time series and brings extra uncertainty to the model.
Thereby, we do not include GANs for comparisons in our numerical experiment.

\section{Numerical Experiment}
\label{numerical experiment}

\subsection{Data and Model Settings}

In this section, we consider the portfolio construction using the component stocks of the Dow Jones Industrial Average index (DJIA) (adopt the stock list on April 6, 2020), and finally select $N=28$ individual stocks with complete daily return records from May 5, 1999 to September 22, 2021. We download corresponding daily adjusted prices 
and calculate percentage change returns as inputs for concerned models. As to the choice of market returns, we consider the S\&P 500 index returns as the single factor for factor-based models.
We divide the whole data set into two parts at February 20, 2013, before which the first $60\%$ is set as the training set, and the last $40\%$ is the testing set for portfolio construction and out-of-sample evaluation. There are totally $3,471$ daily records for training and $2,163$ out-of-sample daily returns for evaluation. 

Note that we aim to examine the model performance on monthly-frequency investment, thus leading to out-of-sample forecasting and portfolio re-balancing $L=103$ times (21 trading days per month, and $2163/21=103$). The portfolio performance of every model is evaluated.
All GRU-based models take in a history window of $T=200$ \emph{daily returns} and forecast the model parameters for future \emph{monthly return}. Now there are still about $3,250$ data points for training, despite that the label or the prediction target is monthly return. The DCC-MM model estimates its parameters with the $165$ monthly returns in the training data ($3471/21\approx 165$). 
For the static SAA, an expanding window technique is adopted, i.e., the $165$ monthly returns are assumed \textit{i.i.d.} and used in the first-time portfolio optimization and then $166,~167,~\dots$ are used in the following. This yields better performance
compared to the rolling window approach in our case.

To prevent overfitting and relieve the concern of tuning parameters, we only use one-layer GRU with $4$ or $6$ hidden nodes and choose the last $20\%$ of the training set as the validation set for early stopping. 
The GRU hidden state dimension is $4$ in the market return fitting step (Algorithm \ref{algorithm_N}), and is $6$ in the individual stock return fitting step (Algorithm \ref{algorithm_i}). These dimensions are twice the output dimensions, which are 2 and 3 respectively. They are not large because we indeed decompose the multi-dimensional problem to many one-dimensional problems. This reduces the difficulty of learning since each asset will have a low-dimensional Attention-GRU.
 
Besides, the following settings are common for both the algorithms: $l_{\text{fix}}=10^{-2}$, $l_{\text{tv}}=10^{-3}$, $N_m=6$, $N_{\text{fix}}=2000$, $N_{\text{tv}}=2000$, and the batch size $S$ is the training set size.
We apply the normalization of zero-mean and unit-variance in data pre-processing.
At last, to ensure the robustness of all GRU-based models, we independently train $B_r=5$ distinct models using different random initialization seeds. Each trained model is then utilized to generate a forecast of the factor model. Subsequently, we aggregate these forecasts by taking their average and obtain an ensemble forecast. The ensemble forecast is then employed in the subsequent CVaR portfolio optimization. This ensemble technique was also adopted in \cite{gu2020empirical} and was a key to improve the performance and the stability of neural networks.

For the large batch size, we know that the standard is using mini-batches. However, based on our experience, it is better to set a large batch size here. When it is common in computer science community to adopt a batch size of 128, 256, 512, and so on (the powers of 2), we found that these small batch sizes did not work well in our study. 
We directly set the batch size to be the training set size, and it works well.
A possible explanation for this is the unstable predictive pattern in financial data. Two mini-batches of data can easily have distinct predictive patterns and the gradient descent may not work as expected. A large batch of samples may include a more stable relationship between the features and the label. 


\subsection{Estimated Time-Invariant Tail Parameters} \label{sec:tail-param}

\begin{table}[t]
	\caption{{The time-invariant tail parameters $\nu_M=\{u_M,v_M\}$ for market returns S\&P 500, and $\nu^M_i=\{u^M_i,v^M_i\}$, $\nu_i=\{u_i,v_i\}$ for individual stock returns in the Dow Jones index, averaged from $B_r=5$ independent trainings of the GF-AGRU model.}}
	\centering
	\footnotesize
		\begin{tabular}{c|cccc||c|cccc}
			\hline
			Ticker & $u_i^M$ & $v_i^M$ & $u_i$ & $v_i$ & Ticker & $u_i^M$ & $v_i^M$ & $u_i$ & $v_i$ \\ \hline
			S\&P 500 & \makecell[c]{$u_M$\\ 1.009} & \makecell[c]{$v_M$\\ 1.630} &  &  &  &  &  &  &  \\ \hline
			AXP & 1.993 & 2.257 & 1.753 & 1.516 &   JPM & 1.370 & 1.568 & 2.000 & 1.288 \\
			XOM & 1.112 & 1.745 & 1.491 & 1.359 &   MCD & 1.927 & 2.606 & 1.691 & 1.756 \\
			AAPL & 1.618 & 1.880 & 1.442 & 1.785 &   MMM & 1.468 & 1.838 & 1.580 & 1.509 \\
			BA & 1.374 & 2.426 & 1.346 & 1.361 &   MRK & 2.509 & 2.692 & 1.462 & 1.721 \\
			CAT & 1.439 & 1.699 & 1.690 & 1.520 &  MSFT & 1.271 & 1.240 & 1.900 & 1.855 \\
			CSCO & 1.531 & 1.743 & 1.618 & 1.547 &   NKE & 1.939 & 2.216 & 1.912 & 1.817 \\
			CVX & 1.309 & 1.743 & 1.423 & 1.464 &    PG & 1.237 & 2.189 & 1.467 & 1.850 \\
			GS & 1.465 & 1.553 & 1.725 & 1.610 &   TRV & 2.660 & 2.842 & 2.032 & 1.494 \\
			HD & 1.737 & 2.033 & 1.401 & 1.254 &   UNH & 1.747 & 2.121 & 1.361 & 1.653 \\
			PFE & 1.551 & 1.851 & 1.756 & 1.421 &   RTX & 1.050 & 1.663 & 1.786 & 1.848 \\
			IBM & 1.358 & 1.404 & 1.961 & 1.944 &    VZ & 2.241 & 2.085 & 1.677 & 1.404 \\
			INTC & 1.479 & 1.680 & 1.333 & 1.701 &   WBA & 1.650 & 2.044 & 1.533 & 1.621 \\
			JNJ & 2.648 & 2.829 & 1.577 & 1.309 &   WMT & 1.912 & 1.841 & 1.493 & 1.403 \\
			KO & 1.855 & 2.296 & 1.290 & 1.501 &   DIS & 1.417 & 1.906 & 1.492 & 1.574 \\
			\hline
	\end{tabular}
	\label{tab:tailparam-seed}
\end{table}

We present the estimated time-invariant tail parameters in the GF-AGRU model in Table \ref{tab:tailparam-seed}. Note that these estimated parameters are the averages from the $B_r=5$ independent trainings using different seeds. The column \emph{Ticker} gives the symbols of the included stocks. As described in the \ref{gx-factor} model, we present the two tail parameters $\nu_M=\{u_M,v_M\}$ for the monthly returns of S$\&$P $500$ index, which serve as the market returns. The right and left tail parameters of the market returns are $1.009$ and $1.630$ respectively, which indicates a heavier tail in the lower loss than in the upper reward. The number $1.009$ on the upper side implies an approximate Gaussian right tail (the number 1 implies an exact Gaussian tail). 

With regards to the individual stocks, the tail parameters are divided into two parts: $\nu_i^M=\{u^M_i,v^M_i\}$ is concerned with the tail sensitivities of the stock to the common market factor; and $\nu_i=\{u_i,v_i\}$ controls the residual tails or the idiosyncratic tails. We observe that $v_i^M$ is relatively larger than $v_i$ for most of these individual stocks, which suggests that the individual left tail heaviness is mainly decided by the extreme loss attributed to the market factor. Another observation is that for most stocks, $v_i^M$ is larger than $u_i^M$ and both are larger than $1$, which indicates that the stocks exhibit widely-existing tail dependence with the market returns (and with each other) and the lower tail dependence (left-tail side) is significantly larger than the upper.

Recall that the closer the tail parameters are to one, the more the distribution tails resemble Gaussian tails, or equally the lighter the tails are. Table \ref{tab:tailparam-seed} shows that most stocks have heavier tails than the normal distribution. And in most cases, the tails are asymmetric on left and right sides. 
For example, a larger $v_i^M$ with a smaller $u_i^M$ implies a heavier left tail (attributed to the market factor) than the right, as observed in many stocks. However, the residual tails show different patterns across different stocks. For example, IBM has similar $u_i$ and $v_i$ around $1.95$, and the residual tails of JPM are obviously asymmetric and right-skewed (upper tail $u_i=2.000$ and lower tail $v_i=1.288$), while the residual tails of MRK are left-skewed (upper tail $u_i=1.462$ and lower tail $v_i=1.721$).

\subsection{Predicted Time-Varying Parameters and Coverage Tests} \label{sec:param-and-coverage}

In addition to the time-invariant tail parameters, we can also take a look at the time-varying parameters $\Theta^t_{\text{GF}}$ predicted by us in the testing/out-of-sample set. They include the time-varying parameters of market returns $\Theta^t_{M}=\{\alpha^t_M,\beta^t_M\}$ and the ones of individual stock returns $\Theta^t_{i}=\{\alpha^t_i,\beta^t_i,\gamma^t_i\}$. The visualizations of their temporal behaviors are in \ref{AppendixA}, in which we show the curve results of S\&P 500 and 10 representative stocks selected in the Dow Jones index. One can find that many stocks show a spike in the year of 2020, which may be attributed to the crash caused by the pandemic.

Because we have forecasted the joint distribution of stock returns in each month in the testing period, we can easily test how well the proposed model predicts the return distribution of each stock. It can partially support that our method leads to a good fit for the data. Focusing on the tail risk estimations, we conduct some coverage tests to backtest the predictions of VaR at 99\%, 95\%, or 90\% level for each stock. The results and analysis are in \ref{AppendixB}.

\subsection{Portfolio Optimization Settings}

Next, we focus on the investment performance of portfolio constructions given by several models with the CVaR objective in Equation \eqref{eqn:CVaR-SAA} and Equation \eqref{eqn:CVaR-empirical}. 
The naive EW diversification will allocate equal weights to all stocks regardless of the past returns. As to the other models, the optimal weights are obtained by solving the CVaR minimization problem with different confidence levels $q$: $q \in \{0.90,0.95,0.99\} $. At each investment date, the number of new samples generated by the forecasted generative factor model (act as the inputs to the CVaR minimization problem) is $n\in\{5000,10000,50000\}$, corresponding to the different values of $q$ respectively. It satisfies $n(1-q)=500$.

Noticing that the overall average monthly return of all stocks is around $2\%$, we accordingly set the target return $R_0$ in CVaR minimization as $R_0\in\{0.01,0.02,0.03\}$ (one of the optimization constraints), denoted as $R_1$, $R_2$, and $R_3$ respectively. Besides, we also consider chasing the realized return obtained by the equally weighted portfolio, producing a time-varying $R_0$ and denoting this target return strategy as $R_{\text{ew}}$, which is supposed to improve the performance of the optimal portfolio as suggested in \citet{kirby2012s} and \citet{hwang2018naive}. It shall be noted that we conduct the simulation and CVaR optimization 10 times and report the average result to alleviate the randomness effect of  sampling. 

\subsubsection{Evaluation Metrics}

The weights $\tilde{w}_t$ obtained by solving the minimization will be used to construct the portfolio at each investment date and obtain a one-month future return $\tilde{w}^\top_tY^t$, where $Y^t$ now is the actual monthly returns of the $N$ stocks. Then we move to the next month (21 trading days) for re-balancing the portfolio, and repeat the process until we obtain $L=103$ dynamic returns, denoted as $\tilde{R}=\{\tilde{R}_1,\dots,\tilde{R}_L\}$. For evaluating the performance of $\tilde{R}$, we measure its average return (AV), standard deviation (SD), and their ratio denoted as information ratio (IR). We annualize these measures and calculate
\begin{equation}
\begin{aligned}
     \text{AV} &= E[\tilde{R}]\times 12,\\
     \text{SD} &= \sqrt{\mathrm{Var}(\tilde{R})\times 12},\\
     \text{IR} &= \text{AV}/\text{SD},
\end{aligned}
\end{equation}
where $E[\cdot]$ is the empirical mean and $\mathrm{Var}(\cdot)$ is the empirical variance. Besides, we are also interested in the tail risk of the portfolio and thus include the \(95\%\) lower-tail CVaR (Expected Shortfall, ES) as a performance measure which is obtained by
\begin{equation}
    \text{ES} = E[-\tilde{R}|-\tilde{R}\geq \text{VaR}_{95\%}(-\tilde{R})]\times 12.
\end{equation}
A small ES is expected for well-performed portfolios. 
Besides, we also calculate the skewness (SK) of $\tilde{R}$ and the maximum drawdown (MD) of the portfolio as performance measures. 

At last, noticing that investors do not share an unilateral risk aversion toward the upper and the lower tails of the returns, therefore we also include a CVaR-based ratio called Stable-Tail Adjusted Return Ratio or STARR (simplified as CR) proposed by \citet{martin2003phi}. CR is defined as
 \begin{equation}
      \text{CR} = \frac{E[\tilde{R}]}{E[-\tilde{R}|-\tilde{R}\geq \text{VaR}_{95\%}(-\tilde{R})]},
 \end{equation}
 where the denominator is the \(95\%\) lower-tail CVaR of the portfolio. Similarly, the Rachev ratio (RR)  given by \citet{biglova2004different} calculates the ratio of upper expected excess and the lower-tail CVaR:
  \begin{equation}
     \text{RR} = \frac{E[\tilde{R}|-\tilde{R}\leq \text{VaR}_{95\%}(-\tilde{R})]}{E[-\tilde{R}|-\tilde{R}\geq \text{VaR}_{95\%}(-\tilde{R})]}.
 \end{equation}
 The STARR ratio (CR) and Rachev ratio (RR) provide better reward-risk measures for skewed and heavy-tailed portfolio returns  (see \citet{farinelli2008beyond}).

\subsection{Discussion on Performance}

\begin{sidewaystable}
	\caption{Investment performance of CVaR portfolio optimization on the Dow Jones stocks with different settings of confidence level $q$ and sample size $n$. Four models are considered for comparisons: EW, static SAA (expanding window), DCC-MM, and the proposed GF-AGRU.
		$R_1$, $R_2$, and $R_3$ represent the target returns of $0.01$, $0.02$, and $0.03$, respectively. $R_{\text{ew}}$ represents the time-varying target return achieved by EW. Given a specific target return, the best performance under each evaluation metric is displayed with bold.} 
	\scriptsize
	\centering
	
	\begin{tabular}{cccccccccccccccc}
		\hline
		\multicolumn{16}{l}{\textbf{Panel A: CVaR Optimization with $q=0.90$, $n=5000$}} \\
		\hline
		Metric & EW & \multicolumn{4}{c}{SAA} && \multicolumn{4}{c}{DCC-MM} && \multicolumn{4}{c}{GF-AGRU}    \\ 
		\cline{3-6} \cline{8-11} \cline{13-16}   
		& &  $R_1$ &$R_2$  &$R_3$    & $R_{\text{ew}} $   &&  
		$R_1$ &$R_2$  &$R_3$    & $R_{\text{ew}} $   && 
		$R_1$ &$R_2$  &$R_3$    & $R_{\text{ew}} $   \\
		\hline
		AV &  0.151 &  0.139 &            0.26 &  \textbf{0.331} &  0.208 &   &           0.167 &   0.284 &  0.313 &           0.254 &   &  \textbf{0.243} &  \textbf{0.311} &           0.318 &  \textbf{0.334} \\
		SD &  0.139 &  0.146 &  \textbf{0.161} &           0.292 &  0.236 &   &  \textbf{0.135} &   0.172 &  0.264 &           0.226 &   &           0.171 &           0.183 &  \textbf{0.238} &   \textbf{0.22} \\
		IR &  1.088 &   0.95 &           1.615 &           1.132 &  0.884 &   &           1.237 &   1.657 &  1.184 &           1.121 &   &   \textbf{1.42} &  \textbf{1.698} &  \textbf{1.337} &  \textbf{1.517} \\
		MD &  0.256 &  0.211 &           0.211 &           0.314 &  0.206 &   &           0.215 &   0.233 &  0.284 &  \textbf{0.202} &   &  \textbf{0.202} &  \textbf{0.196} &  \textbf{0.236} &  \textbf{0.202} \\
		ES &  1.133 &  1.059 &   \textbf{1.02} &           1.832 &  1.379 &   &            1.04 &   1.158 &   1.72 &           1.347 &   &    \textbf{1.0} &           1.053 &  \textbf{1.491} &  \textbf{1.155} \\
		SK & -1.610 &  0.009 &          -0.147 &           0.241 &  1.294 &   &          -0.747 &  -0.345 &  0.062 &           1.251 &   &  \textbf{0.213} &   \textbf{0.21} &  \textbf{0.277} &  \textbf{1.314} \\
		CR &  0.133 &  0.131 &           0.254 &            0.18 &  0.151 &   &           0.161 &   0.246 &  0.182 &           0.188 &   &  \textbf{0.243} &  \textbf{0.295} &  \textbf{0.213} &  \textbf{0.289} \\
		RR &  0.191 &  0.189 &           0.318 &           0.241 &   0.21 &   &            0.22 &   0.309 &  0.242 &           0.249 &   &  \textbf{0.307} &  \textbf{0.361} &  \textbf{0.275} &  \textbf{0.355} \\

	\end{tabular}
	
	\begin{tabular}{cccccccccccccccc}
		\hline
		\multicolumn{16}{l}{\textbf{Panel B: CVaR Optimization with $q=0.95$, $n=10000$}} \\
		\hline
		Metric & EW & \multicolumn{4}{c}{SAA} && \multicolumn{4}{c}{DCC-MM} && \multicolumn{4}{c}{GF-AGRU}    \\ 
		\cline{3-6} \cline{8-11} \cline{13-16}     
		& &  $R_1$ &$R_2$  &$R_3$    & $R_{\text{ew}} $   &&      
		$R_1$ &$R_2$  &$R_3$    & $R_{\text{ew}} $  && 
		$R_1$ &$R_2$  &$R_3$    & $R_{\text{ew}} $    \\
		\hline
		AV &  0.151 &  0.151 &           0.257 &  \textbf{0.331} &           0.215 &   &           0.167 &   0.282 &  0.311 &         0.253 &   &  \textbf{0.242} &  \textbf{0.318} &           0.323 &  \textbf{0.328} \\
		SD &  0.139 &  0.142 &  \textbf{0.161} &           0.292 &           0.235 &   &  \textbf{0.135} &   0.172 &  0.263 &         0.227 &   &           0.167 &           0.183 &  \textbf{0.238} &  \textbf{0.222} \\
		IR &  1.088 &  1.061 &           1.595 &           1.132 &           0.915 &   &           1.239 &   1.644 &  1.181 &         1.118 &   &  \textbf{1.452} &  \textbf{1.739} &  \textbf{1.357} &   \textbf{1.48} \\
		MD &  0.256 &  0.204 &           0.214 &           0.314 &           0.211 &   &           0.217 &   0.237 &  0.278 &  \textbf{0.2} &   &  \textbf{0.202} &  \textbf{0.196} &  \textbf{0.234} &           0.202 \\
		ES &  1.133 &  1.047 &  \textbf{1.005} &           1.832 &           1.335 &   &           1.041 &   1.168 &  1.692 &         1.367 &   &    \textbf{1.0} &           1.053 &  \textbf{1.483} &  \textbf{1.159} \\
		SK & -1.610 &  0.019 &          -0.141 &           0.241 &  \textbf{1.318} &   &          -0.809 &  -0.372 &  0.109 &         1.234 &   &  \textbf{0.235} &  \textbf{0.162} &  \textbf{0.245} &           1.276 \\
		CR &  0.133 &  0.144 &           0.256 &            0.18 &           0.161 &   &           0.161 &   0.242 &  0.184 &         0.185 &   &  \textbf{0.242} &  \textbf{0.302} &  \textbf{0.218} &  \textbf{0.283} \\
		RR &  0.191 &  0.202 &            0.32 &           0.241 &            0.22 &   &            0.22 &   0.305 &  0.244 &         0.246 &   &  \textbf{0.306} &  \textbf{0.368} &   \textbf{0.28} &  \textbf{0.349} \\

	\end{tabular}
	
	\begin{tabular}{cccccccccccccccc}
		\hline
		\multicolumn{16}{l}{\textbf{Panel C: CVaR Optimization with $q=0.99$, $n=50000$}} \\
		\hline
		Metric & EW & \multicolumn{4}{c}{SAA} && \multicolumn{4}{c}{DCC-MM} && \multicolumn{4}{c}{GF-AGRU}    \\ 
		\cline{3-6} \cline{8-11} \cline{13-16}     
		& &  $R_1$ &$R_2$  &$R_3$    & $R_{\text{ew}} $  &&      
		$R_1$ &$R_2$  &$R_3$    & $R_{\text{ew}} $  && 
		$R_1$ &$R_2$  &$R_3$    & $R_{\text{ew}} $  \\
		\hline
		AV &  0.151 &   0.133 &           0.245 &  \textbf{0.331} &  0.184 &   &           0.168 &   0.286 &  0.312 &         0.252 &   &  \textbf{0.218} &  \textbf{0.317} &           0.324 &  \textbf{0.309} \\
		SD &  0.139 &   0.148 &  \textbf{0.168} &           0.292 &  0.242 &   &  \textbf{0.135} &   0.172 &  0.263 &         0.227 &   &           0.159 &           0.183 &  \textbf{0.238} &   \textbf{0.22} \\
		IR &  1.088 &   0.901 &           1.462 &           1.132 &  0.761 &   &           1.243 &   1.662 &  1.189 &         1.111 &   &  \textbf{1.372} &  \textbf{1.737} &   \textbf{1.36} &  \textbf{1.402} \\
		MD &  0.256 &   0.249 &           0.214 &           0.314 &  0.276 &   &           0.215 &   0.235 &  0.279 &  \textbf{0.2} &   &  \textbf{0.202} &  \textbf{0.196} &  \textbf{0.234} &           0.202 \\
		ES &  1.133 &   1.189 &           1.136 &           1.832 &  1.485 &   &           1.041 &   1.152 &  1.693 &         1.364 &   &  \textbf{0.971} &   \textbf{1.05} &  \textbf{1.478} &  \textbf{1.158} \\
		SK & -1.610 &  -0.451 &          -0.147 &           0.241 &  1.109 &   &          -0.772 &  -0.345 &   0.11 &         1.239 &   &  \textbf{0.216} &  \textbf{0.173} &   \textbf{0.26} &  \textbf{1.359} \\
		CR &  0.133 &   0.112 &           0.216 &            0.18 &  0.124 &   &           0.161 &   0.248 &  0.185 &         0.185 &   &  \textbf{0.225} &  \textbf{0.302} &  \textbf{0.219} &  \textbf{0.267} \\
		RR &  0.191 &   0.169 &           0.278 &           0.241 &  0.181 &   &           0.221 &   0.312 &  0.245 &         0.245 &   &  \textbf{0.287} &  \textbf{0.369} &  \textbf{0.281} &  \textbf{0.331} \\

		\hline
		
	\end{tabular}
	
	
	\label{benchmark2}
\end{sidewaystable}

Table \ref{benchmark2} exhibits the portfolio performance given by four models: EW, static SAA (expanding window), DCC-MM, and the proposed GF-AGRU, under different settings of confidence level $q$ and sample size $n$. Four different target return strategies are considered.  We find that EW leads to a relatively small average return ($\text{AV}=0.151$) and a low information ratio ($\text{IR}=1.088$). Meanwhile, the skewness ($\text{SK}=-1.61$) is significantly negative and implies a tendency for a heavy loss of the portfolio. Unlike the robust performance of the EW strategy as illustrated in \citet{demiguel2009optimal}, our experiment shows that the more advanced models can provide better investment performance, by solving a portfolio optimization problem with the CVaR objective (some of them forecast and simulate from the joint distribution of stock returns).

As we can see from the table, when comparing the static SAA with the EW method, SAA demonstrates some outperformance. For example, when the target return is $R_2 = 0.02$ and the confidence level is $q=0.90$, the SAA model achieves an Information Ratio (IR) of $1.615$, which is notably higher than the IR of EW. Although it is still smaller than the IR of $1.657$ achieved by the DCC-MM model, it should be noted that SAA yields smaller maximum drawdown (MD) and expected shortfall (ES) and slightly higher skewness (SK), CR, and RR compared to DCC-MM under $R_2$. However, for other target returns, SAA falls short in comparisons to DCC-MM under most of the metrics and even performs worse than the EW strategy under $R_1$ and $R_{\text{ew}}$. 
As to the DCC-MM model, we observe that it achieves a high information ratio ($\text{IR}=1.657$) when the target return is $R_2=0.02$. However, DCC-MM fails to persist the good performance when we optimize the CVaR objective with other target returns. Especially with $R_{\text{ew}}$, the information ratio ($\text{IR}= 1.121$) is close to that given by EW. Overall, DCC-MM performs slightly better than the static SAA, and the static SAA only outperforms EW under $R_2$.

When comparing these benchmarks with the proposed GF-AGRU model, it becomes evident that GF-AGRU consistently outperforms them in terms of most metrics across various settings such as the confidence levels of CVaR and the target returns.
Notably, the introduction of the dynamic \ref{gx-factor} model with tail-property considerations leads to a significant improvement in portfolio performance. The GF-AGRU model gives the largest IR for most target returns while also exhibiting superiority in the control of tail risk with small MD \& ES and large SK. Considering the ratios of reward and tail risk, the STARR ratio (CR) and Rachev ratio (RR), we notice that the GF-AGRU model shows consistent outperformance over other models, regardless of the confidence levels of CVaR and the target return constraints. We ascribe the outstanding performance of the GF-AGRU model to the detailed depiction of the tail properties of stock returns, the succinct generative model structure, the powerful attention mechanism combined with GRU, and the well-designed training algorithm which decomposes a difficult multi-dimensional learning task into many separate one-dimensional tasks.

\subsection{Ablation Study}

\begin{sidewaystable}
\caption{Inspect the importance of each part in the GF-AGRU model (ablation study), by exhibiting the investment performance of CVaR portfolio optimization on the Dow Jones stocks with different settings of confidence level $q$ and sample size $n$. Four models are considered for comparisons: EW,  Naive-AGRU, GF-GRU, and the proposed GF-AGRU.
$R_1$, $R_2$, and $R_3$ represent the target returns of $0.01$, $0.02$, and $0.03$, respectively. $R_{\text{ew}}$ represents the time-varying target return achieved by EW. Given a specific target return, the best performance under each evaluation metric is displayed with bold.} 
\scriptsize
\centering

\begin{tabular}{cccccccccccccccc}
\hline
\multicolumn{16}{l}{\textbf{Panel A: CVaR Optimization with $q=0.90$, $n=5000$}} \\
\hline
Metric & EW & \multicolumn{4}{c}{Naive-AGRU} && \multicolumn{4}{c}{GF-GRU} && \multicolumn{4}{c}{GF-AGRU}    \\ 
\cline{3-6} \cline{8-11} \cline{13-16}   
  & &  $R_1$ &$R_2$  &$R_3$    & $R_{\text{ew}} $   &&  
       $R_1$ &$R_2$  &$R_3$    & $R_{\text{ew}} $   && 
    $R_1$ &$R_2$  &$R_3$    & $R_{\text{ew}} $   \\
    \hline
 AV &  0.151 &           0.165 &           0.231 &           0.285 &           0.242 &   &    0.16 &   0.233 &  0.278 &  0.249 &   &  \textbf{0.243} &  \textbf{0.311} &  \textbf{0.318} &  \textbf{0.334} \\
 SD &  0.139 &  \textbf{0.132} &  \textbf{0.149} &  \textbf{0.206} &  \textbf{0.216} &   &   0.182 &   0.195 &  0.229 &  0.237 &   &           0.171 &           0.183 &           0.238 &            0.22 \\
 IR &  1.088 &           1.253 &           1.553 &  \textbf{1.381} &           1.118 &   &   0.876 &   1.196 &  1.214 &  1.053 &   &   \textbf{1.42} &  \textbf{1.698} &           1.337 &  \textbf{1.517} \\
 MD &  0.256 &           0.209 &           0.212 &  \textbf{0.218} &  \textbf{0.195} &   &   0.273 &   0.276 &  0.248 &  0.251 &   &  \textbf{0.202} &  \textbf{0.196} &           0.236 &           0.202 \\
 ES &  1.133 &           1.052 &           1.102 &  \textbf{1.379} &           1.224 &   &    1.26 &   1.362 &  1.546 &  1.385 &   &    \textbf{1.0} &  \textbf{1.053} &           1.491 &  \textbf{1.155} \\
 SK & -1.610 &          -0.571 &          -0.318 &           -0.05 &  \textbf{1.532} &   &  -0.685 &  -0.521 &  0.089 &  0.845 &   &  \textbf{0.213} &   \textbf{0.21} &  \textbf{0.277} &           1.314 \\
 CR &  0.133 &           0.157 &            0.21 &           0.207 &           0.198 &   &   0.127 &   0.171 &   0.18 &   0.18 &   &  \textbf{0.243} &  \textbf{0.295} &  \textbf{0.213} &  \textbf{0.289} \\
 RR &  0.191 &           0.216 &           0.272 &           0.268 &           0.259 &   &   0.184 &   0.231 &   0.24 &   0.24 &   &  \textbf{0.307} &  \textbf{0.361} &  \textbf{0.275} &  \textbf{0.355} \\

\end{tabular}

\begin{tabular}{cccccccccccccccc}
\hline
\multicolumn{16}{l}{\textbf{Panel B: CVaR Optimization with $q=0.95$, $n=10000$}} \\
\hline
Metric & EW & \multicolumn{4}{c}{Naive-AGRU} && \multicolumn{4}{c}{GF-GRU} && \multicolumn{4}{c}{GF-AGRU}    \\ 
\cline{3-6} \cline{8-11} \cline{13-16}     
  & &  $R_1$ &$R_2$  &$R_3$    & $R_{\text{ew}} $   &&      
       $R_1$ &$R_2$  &$R_3$    & $R_{\text{ew}} $  && 
    $R_1$ &$R_2$  &$R_3$    & $R_{\text{ew}} $    \\
    \hline
AV &  0.151 &           0.165 &           0.229 &           0.283 &           0.241 &   &   0.169 &   0.235 &  0.278 &  0.242 &   &  \textbf{0.242} &  \textbf{0.318} &  \textbf{0.323} &  \textbf{0.328} \\
 SD &  0.139 &  \textbf{0.132} &  \textbf{0.149} &  \textbf{0.205} &  \textbf{0.216} &   &   0.183 &   0.196 &  0.229 &  0.238 &   &           0.167 &           0.183 &           0.238 &           0.222 \\
 IR &  1.088 &           1.246 &           1.539 &  \textbf{1.377} &           1.115 &   &   0.921 &   1.198 &  1.213 &  1.017 &   &  \textbf{1.452} &  \textbf{1.739} &           1.357 &   \textbf{1.48} \\
 MD &  0.256 &           0.212 &           0.218 &   \textbf{0.22} &  \textbf{0.195} &   &   0.268 &   0.279 &  0.248 &  0.241 &   &  \textbf{0.202} &  \textbf{0.196} &           0.234 &           0.202 \\
 ES &  1.133 &           1.052 &           1.107 &  \textbf{1.372} &           1.231 &   &   1.233 &   1.325 &  1.526 &  1.385 &   &    \textbf{1.0} &  \textbf{1.053} &           1.483 &  \textbf{1.159} \\
 SK & -1.610 &          -0.621 &          -0.384 &           -0.05 &  \textbf{1.532} &   &  -0.578 &  -0.403 &  0.135 &  0.853 &   &  \textbf{0.235} &  \textbf{0.162} &  \textbf{0.245} &           1.276 \\
 CR &  0.133 &           0.157 &           0.207 &           0.206 &           0.196 &   &   0.137 &   0.177 &  0.182 &  0.174 &   &  \textbf{0.242} &  \textbf{0.302} &  \textbf{0.218} &  \textbf{0.283} \\
 RR &  0.191 &           0.216 &           0.269 &           0.268 &           0.257 &   &   0.195 &   0.237 &  0.242 &  0.234 &   &  \textbf{0.306} &  \textbf{0.368} &   \textbf{0.28} &  \textbf{0.349} \\

\end{tabular}
  
\begin{tabular}{cccccccccccccccc}
\hline
\multicolumn{16}{l}{\textbf{Panel C: CVaR Optimization with $q=0.99$, $n=50000$}} \\
\hline
Metric & EW & \multicolumn{4}{c}{Naive-AGRU} && \multicolumn{4}{c}{GF-GRU} && \multicolumn{4}{c}{GF-AGRU}    \\ 
\cline{3-6} \cline{8-11} \cline{13-16}     
  & &  $R_1$ &$R_2$  &$R_3$    & $R_{\text{ew}} $  &&      
       $R_1$ &$R_2$  &$R_3$    & $R_{\text{ew}} $  && 
    $R_1$ &$R_2$  &$R_3$    & $R_{\text{ew}} $  \\
    \hline
AV &  0.151 &           0.165 &           0.232 &           0.285 &            0.24 &   &   0.179 &   0.243 &  0.282 &  0.265 &   &  \textbf{0.218} &  \textbf{0.317} &  \textbf{0.324} &  \textbf{0.309} \\
 SD &  0.139 &  \textbf{0.132} &  \textbf{0.149} &  \textbf{0.205} &  \textbf{0.216} &   &   0.178 &   0.197 &   0.23 &  0.237 &   &           0.159 &           0.183 &           0.238 &            0.22 \\
 IR &  1.088 &            1.25 &           1.562 &   \textbf{1.39} &           1.107 &   &   1.001 &   1.239 &   1.23 &  1.119 &   &  \textbf{1.372} &  \textbf{1.737} &            1.36 &  \textbf{1.402} \\
 MD &  0.256 &           0.212 &           0.216 &   \textbf{0.22} &  \textbf{0.194} &   &   0.268 &   0.278 &  0.245 &  0.241 &   &  \textbf{0.202} &  \textbf{0.196} &           0.234 &           0.202 \\
 ES &  1.133 &           1.057 &           1.095 &  \textbf{1.363} &           1.232 &   &   1.199 &   1.303 &  1.508 &   1.37 &   &  \textbf{0.971} &   \textbf{1.05} &           1.478 &  \textbf{1.158} \\
 SK & -1.610 &           -0.61 &           -0.35 &          -0.022 &  \textbf{1.536} &   &  -0.658 &  -0.398 &  0.166 &   0.83 &   &  \textbf{0.216} &  \textbf{0.173} &   \textbf{0.26} &           1.359 \\
 CR &  0.133 &           0.157 &           0.212 &            0.21 &           0.195 &   &   0.149 &   0.187 &  0.187 &  0.193 &   &  \textbf{0.225} &  \textbf{0.302} &  \textbf{0.219} &  \textbf{0.267} \\
 RR &  0.191 &           0.216 &           0.274 &           0.271 &           0.255 &   &   0.208 &   0.247 &  0.248 &  0.254 &   &  \textbf{0.287} &  \textbf{0.369} &  \textbf{0.281} &  \textbf{0.331} \\

\hline

\end{tabular}
 

\label{result-dow-attn-inspect-seed}
\end{sidewaystable}

The GF-AGRU model aggregates the merits of several components, i.e., the generative factor model, the heavy-tail properties incorporated, the GRU network, and the attention mechanism. An inquiry that arises naturally is the extent to which each component contributes to the overall performance. Therefore, we contemplate the exclusion of a certain component and conduct a thorough examination of the individual impact.  Note that the model endeavors to learn the joint distribution of multi-dimensional stock returns with the generative factor model and makes forecasts with a sequential learning network. Therefore, it is essential to retain the factor structure and the GRU network, as these components play fundamental roles in the modeling. The aspects that require specific examination are the heavy-tail properties incorporated and the attention mechanism, which introduce flexibility and enhance the model's capability. These components can be considered as the focus of investigation here.


In Section \ref{sec-naive-factor}, we have introduced an alternative approach, which can be obtained by removing the heavy-tail part from the GF-AGRU model. This approach, referred to as the Naive-AGRU model, solely involves fitting a naive factor model without incorporating the $g(\cdot)$ function described in Equation (\ref{gx-func}). On the other hand, in the case that the attention layer is omitted, we maintain the heavy-tail part but directly map the final hidden states of  GRU  to the forecasts of  model parameters  with a dense linear layer. This variant is denoted as the GF-GRU model, distinguishing it from the GF-AGRU model. 
For all the three models, similarly, we employ separate training procedures using $B_r=5$ distinct seeds and take the averaged forecasts. Subsequently, we apply mean-CVaR optimization and evaluate the investment performance of the three models.

The comparisons of the Naive-AGRU, GF-GRU, and GF-AGRU models in Table \ref{result-dow-attn-inspect-seed} demonstrate that the GF-AGRU model consistently outperforms the other two across various metrics. 
An exception is that the portfolios constructed by Naive-AGRU give the smallest standard deviation (SD) and sometimes the smallest maximum drawdown (MD). But their information ratio (IR), CR, and RR are not satisfactory. This can be interpreted that Naive-AGRU adopts Gaussian distribution assumption for stock returns, and hence minimizing CVaR of the portfolio is equivalent to minimizing its variance (given an expected return constraint). However, the gap between the real data and Gaussian distribution makes the estimations of other metrics not accurate, such as AV, ES, and SK, resulting in relatively low IR, CR, and RR.

These results highlight the significance of the heavy-tail properties incorporated and the attention mechanism, further confirming the superiority of our proposed model. It is surprising that the Naive-AGRU model yields a more effective investment strategy compared to the GF-GRU model that solely incorporates heavy-tail properties without the attention mechanism. Sometimes GF-GRU even performs worse than EW. This suggests that the pure GRU network may struggle to capture the complex patterns of dynamic stock returns, if not powered by the attention mechanism.

\subsection{The Cumulative Return Curves}




\begin{figure}[t]
	\centering
	\centerline{\includegraphics[width=1.2\linewidth]{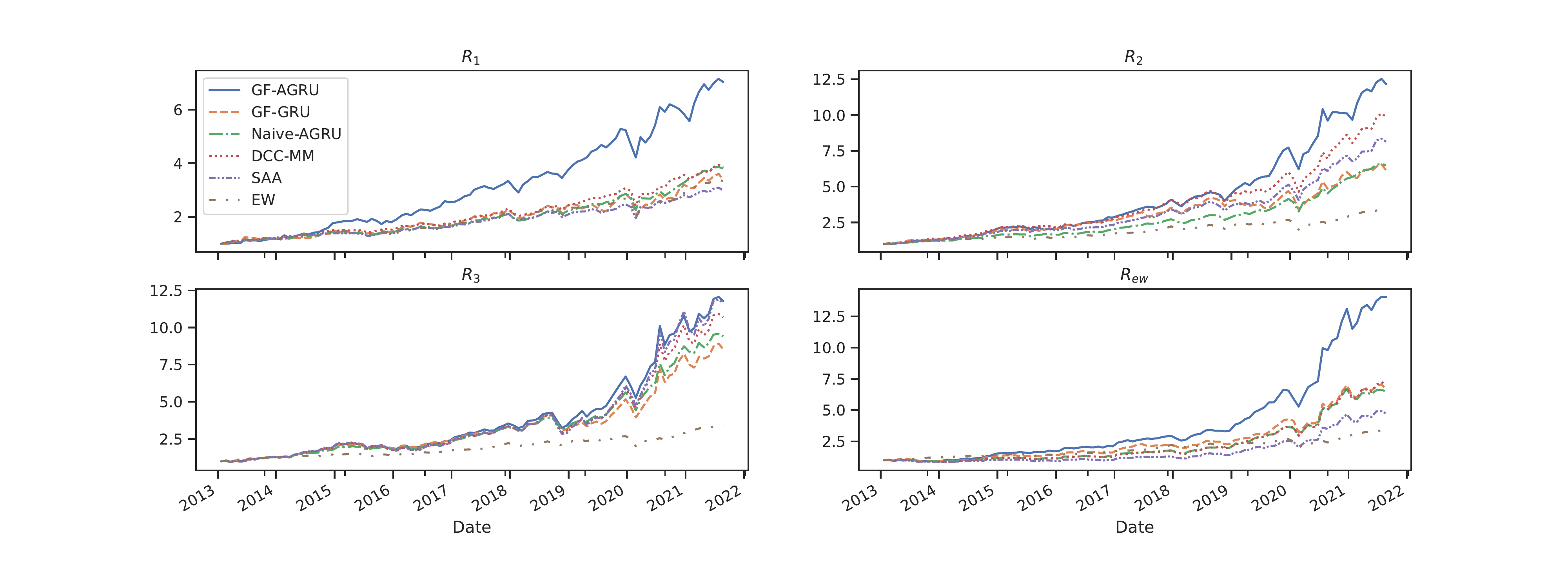}}
	\vspace{-1.5em}
	\caption{The cumulative return curves of the portfolios given by six different models with confidence level $q=0.90$ in CVaR optimization. The four subfigures correspond to four target return constraints respectively, i.e., $R_1=0.01$, $R_2=0.02$, $R_3=0.03$, and $R_{\text{ew}}$.}
	\label{fig:cumq1-new}
\end{figure}

\begin{figure}[t]
	\centering
	\centerline{\includegraphics[width=1.2\linewidth]{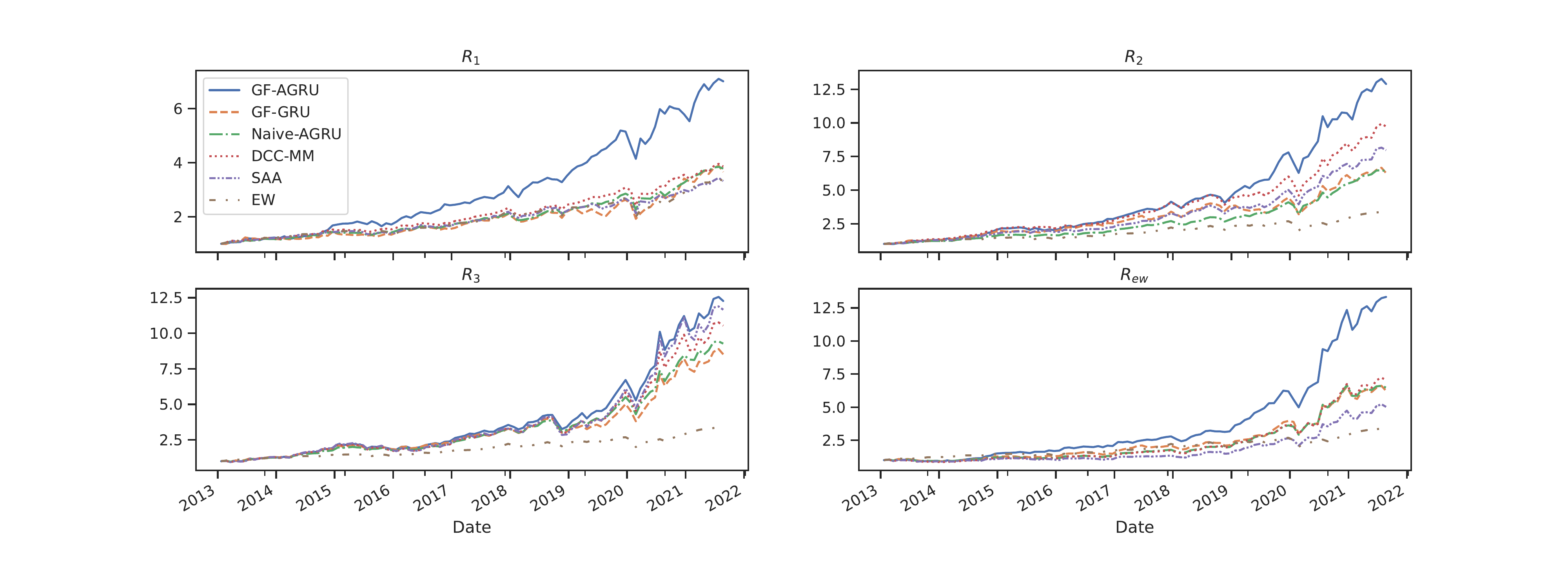}}
	\vspace{-1.5em}
	\caption{The cumulative return curves of the portfolios given by six different models with confidence level $q=0.95$ in CVaR optimization. The four subfigures correspond to four target return constraints respectively, i.e., $R_1=0.01$, $R_2=0.02$, $R_3=0.03$, and $R_{\text{ew}}$.}
	\label{fig:cumq2-new}
\end{figure}

\begin{figure}[t]
	\centering
	\centerline{\includegraphics[width=1.2\linewidth]{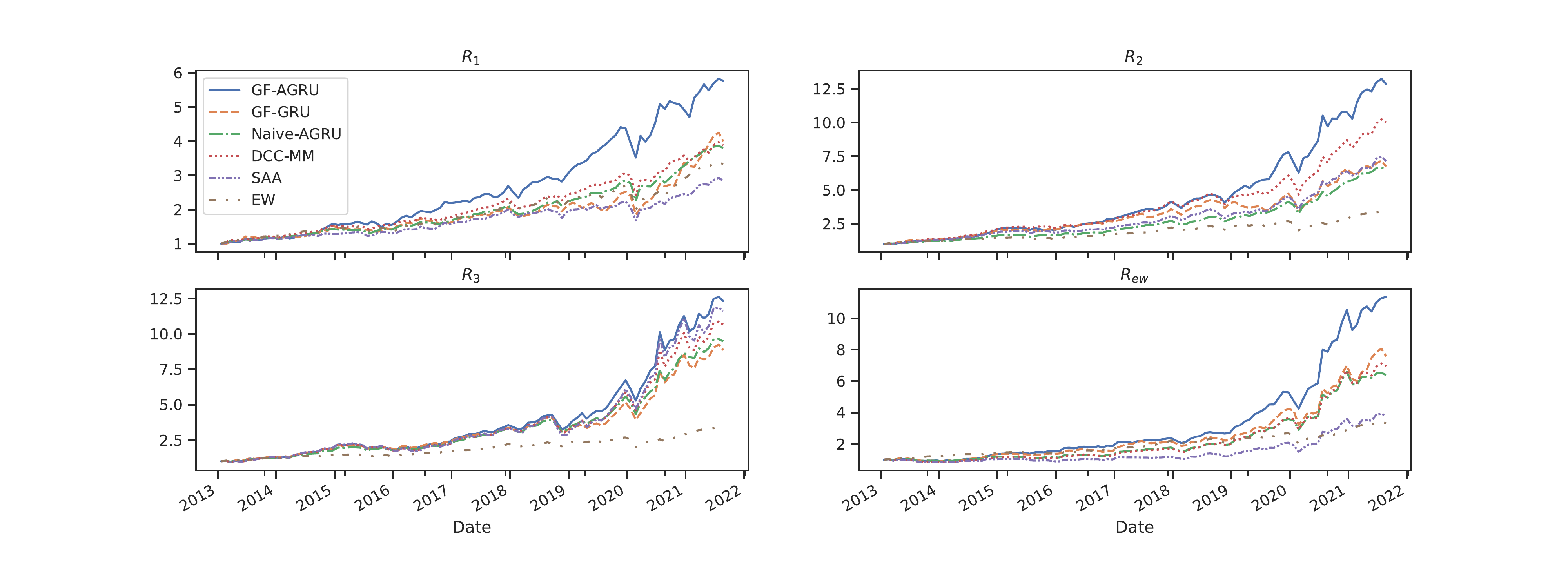}}
	\vspace{-1.5em}
	\caption{The cumulative return curves of the portfolios given by six different models with confidence level $q=0.99$ in CVaR optimization. The four subfigures correspond to four target return constraints respectively, i.e., $R_1=0.01$, $R_2=0.02$, $R_3=0.03$, and $R_{\text{ew}}$.}
	\label{fig:cumq3-new}
\end{figure}

In Figure \ref{fig:cumq1-new}, \ref{fig:cumq2-new}, and \ref{fig:cumq3-new}, we exhibit the cumulative return curves of the portfolios given by the six models considered above: the three benchmark models (EW, SAA, and DCC-MM), the GF-AGRU model, and its two variants (Naive-AGRU and GF-GRU). 
These figures separately show the portfolio performance of six models with different confidence levels in CVaR optimization.
For example, Figure \ref{fig:cumq1-new} plots the results with confidence level $q=0.90$ under various target return constraints, i.e., $R_1=0.01$, $R_2=0.02$, $R_3=0.03$, and $R_{\text{ew}}$. The top-left subfigure shows the cumulative return curves of the portfolios targeting at $R_1=0.01$, from which we observe that the GF-AGRU model remarkably outperforms others and reaches a high profit. It is followed by the DCC-MM model, but its performance closely aligns with the other four models. 

The superiority of GF-AGRU persists with other return constraints (other subfigures in Figure \ref{fig:cumq1-new}) and other confidence levels (Figure \ref{fig:cumq2-new} and \ref{fig:cumq3-new}). 
The static SAA surpasses EW but shows no strength over the DCC-MM model and our model. As to the two variants, i.e.,  Naive-AGRU and GF-GRU, they face challenges in competing with the DCC-MM model. 
We also notice that different target return constraints do significantly influence the portfolio performance, but the change of confidence level $q$ does not make much diffidence to the outcome. 

We observe that the gap between GF-AGRU and other models seems to narrow with target return $R_2$ or $R_3$. However, this does not necessarily lead to the conclusion that their performance are close, because we have many other evaluation metrics as illustrated in Table \ref{benchmark2} and \ref{result-dow-attn-inspect-seed}.
The decreasing gap represents that the average portfolio returns given by the other models are becoming better. This is because the target return constraint increases from $R_1$ to $R_3$. However, our model performs well under various target return constraints. Even under the time varying $R_{\text{ew}}$, our model performs much better than others.
We guess that it is because the stock market becomes better in the testing years and only our approach can benefit from this (other models need to carefully specify a target return constraint).

\subsection{Robustness of Our Model}

From deep learning perspective, it is intrinsic to assess the robustness of our model across various settings or under some inevitable randomness, and demonstrate the consistent advantages offered by our proposed approach. In concrete, we mainly consider three cases of robustness analysis in line with the threads of our numerical experiment, covering the model hyper-parameter specification, the randomness in training the neural network, and the randomness of simulation in CVaR optimization. 
In \ref{AppendixC}, we comprehensively analyze the model robustness in these aspects. The results reveal that our model is robust enough and gives consistent good performance.

\section{Conclusion}
\label{conclusion}

We design a dynamic generative factor model with Attention-GRU neural network to dynamically learn and forecast the model parameters, hence can forecast the joint distribution of multivariate stock returns. 
The factor structure alleviates the concerns about the curse of dimensionality by compressing the high-dimensional joint distribution into a succinct representation, and asymmetric heavy-tail properties are incorporated. 
Then we use many one-layer GRU networks combined with the attention mechanism to predict the parameters of the factor model, including the time-invariant tail parameters. We propose a two-step iterative algorithm to train the proposed GF-AGRU model, which decomposes the difficult multi-dimensional learning problem into many separate one-dimensional learning problems. 

In the numerical experiment, we construct portfolios using the components of the Dow Jones index. The GF-AGRU model learns from the past daily returns and predicts the parameters (thus the joint distribution) of the future monthly returns. We simulate new samples using the generative model learned at each investment date and use the samples to solve the CVaR portfolio optimization problem. Compared with three benchmark models and two GF-AGRU variants, the portfolios constructed by our model can provide higher reward-risk ratios and smaller tail risks. This superiority is consistent with respect to different target return constraints and different confidence levels in CVaR optimization. The superiority also shows adequate robustness. Finally, we find that the attention mechanism and the heavy-tail properties are two key features that make the approach successful.

\appendix

\section{Predicted Time-Varying Parameters $\Theta^t_{\text{GF}}$}
\label{AppendixA}
\addcontentsline{toc}{section}{Appendices}
\renewcommand{\thesubsection}{\Alph{subsection}}

\begin{figure}
	\centering
	\centerline{\includegraphics[width=1\linewidth]{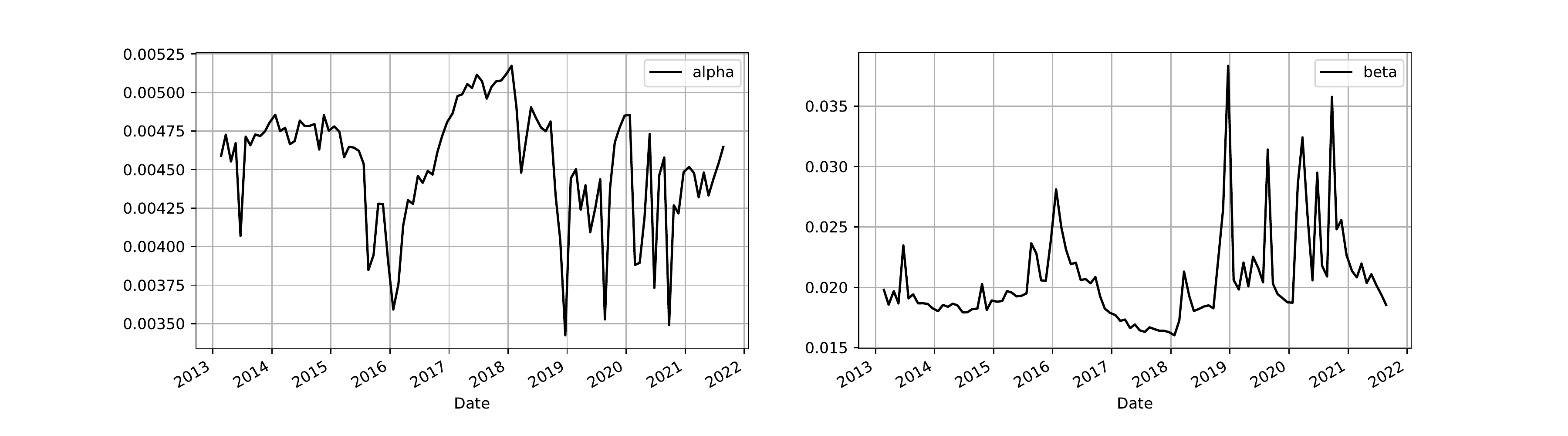}}
	\caption{Predicted time-varying parameters of the market returns (S\&P 500 index). The left shows the curve of predicted $\alpha_M^t$ and the right shows the curve of predicted $\beta_M^t$.}
	\label{fig:param-fig-market}
\end{figure}

\begin{figure}
	\begin{center}
		\begin{subfigure}{1.0\textwidth}      \centerline{\includegraphics[width=1\linewidth]{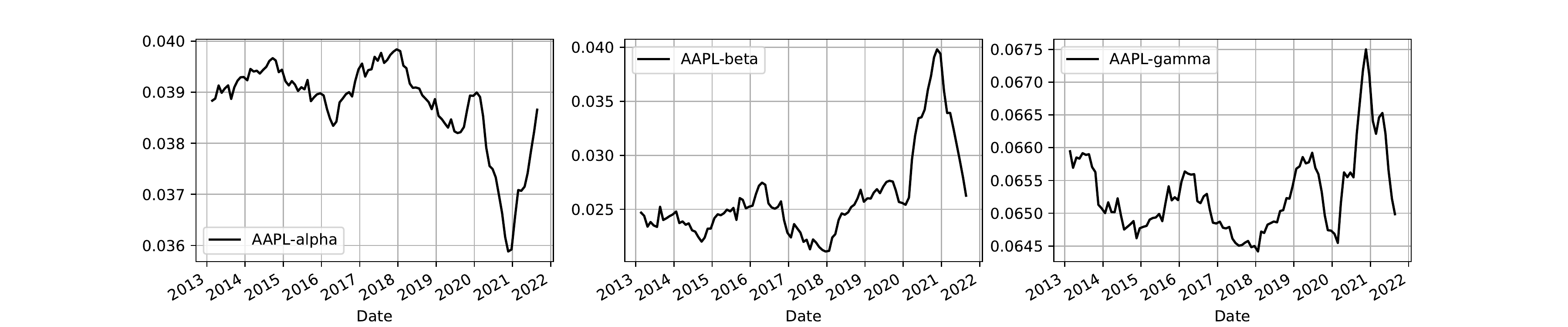}}
		\end{subfigure}
		\begin{subfigure}{1.0\textwidth}           \centerline{\includegraphics[width=1\linewidth]{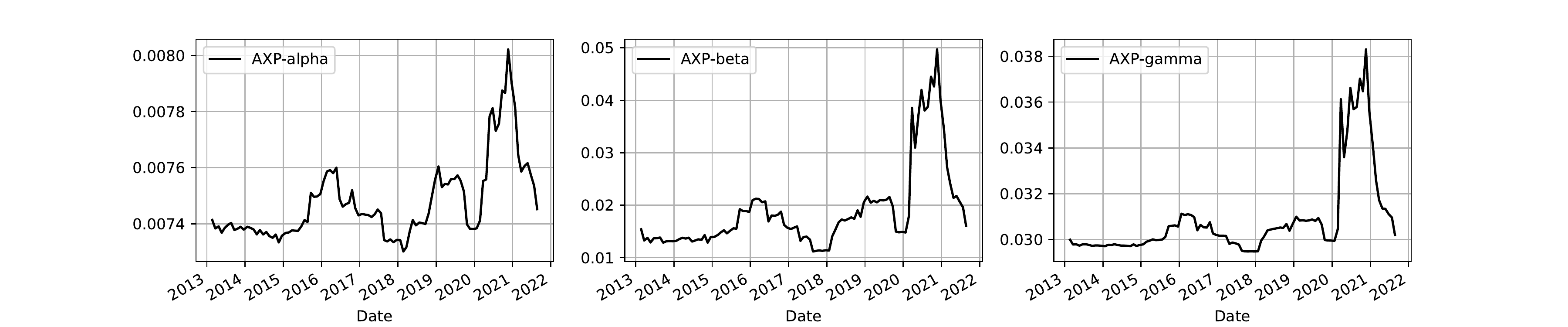}}
		\end{subfigure}
		\begin{subfigure}{1.0\textwidth}           \centerline{\includegraphics[width=1\linewidth]{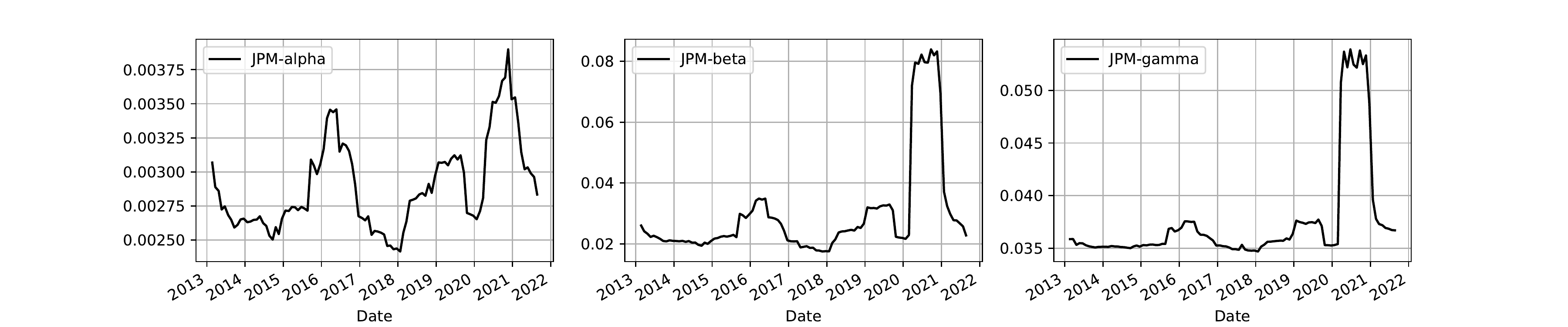}}
		\end{subfigure}
		\begin{subfigure}{1.0\textwidth}          \centerline{\includegraphics[width=1\linewidth]{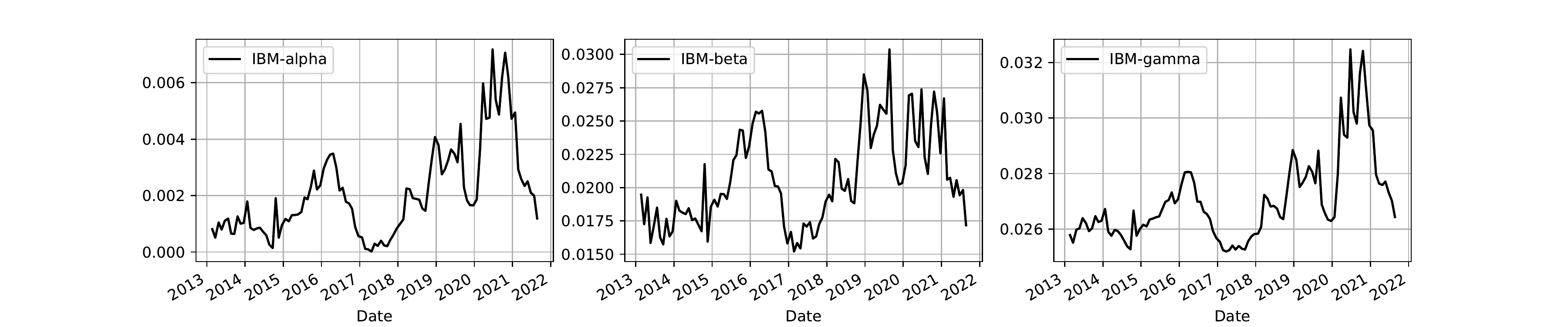}}
		\end{subfigure}
		\begin{subfigure}{1.0\textwidth}         \centerline{\includegraphics[width=1\linewidth]{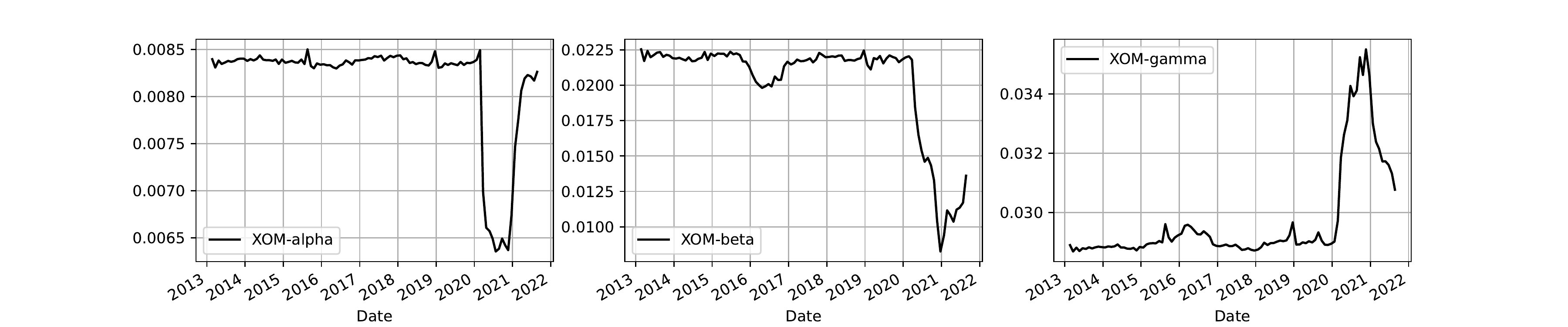}}
		\end{subfigure}
	\end{center}
	\caption{Predicted time-varying parameters of individual stock returns: AAPL, AXP, JPM, IBM, and XOM. The three curves in each row present the dynamics of $\alpha_i^t$ (left), $\beta_i^t$ (middle), and $\gamma_i^t$ (right), predicted by our GF-AGRU model.}
	\label{fig:param-fig-5stocks-1}
\end{figure}

\begin{figure}
	\begin{center}
		\begin{subfigure}{1.0\textwidth}          \centerline{\includegraphics[width=1\linewidth]{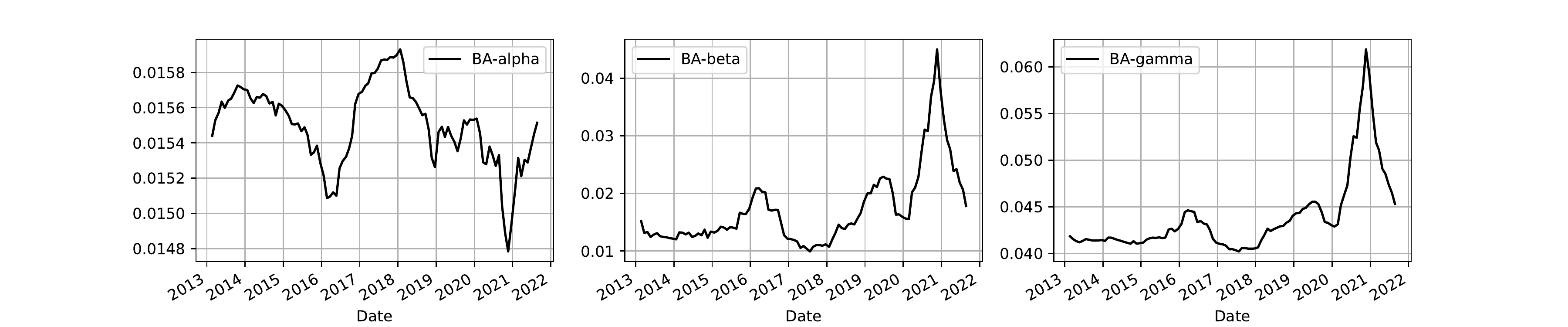}}
		\end{subfigure}
		\begin{subfigure}{1.0\textwidth}          \centerline{\includegraphics[width=1\linewidth]{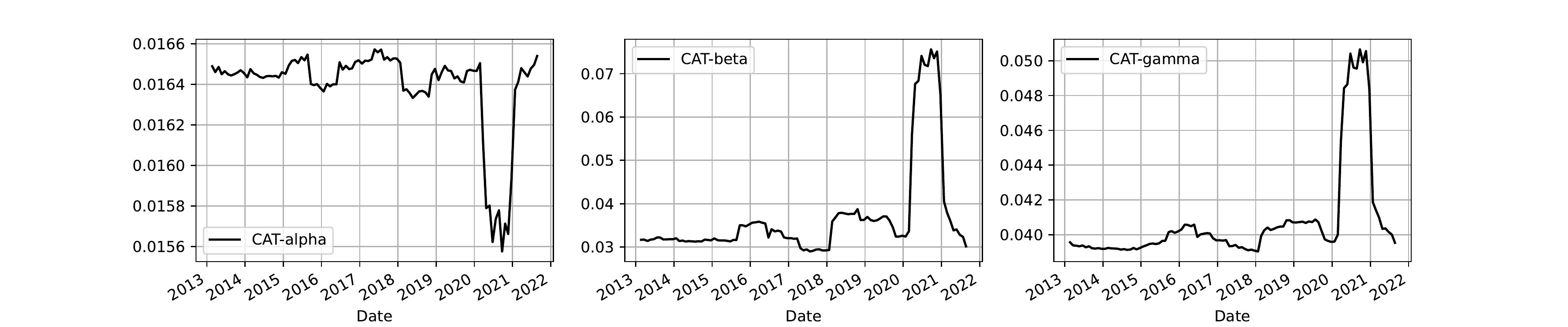}}
		\end{subfigure}
		\begin{subfigure}{1.0\textwidth}          \centerline{\includegraphics[width=1\linewidth]{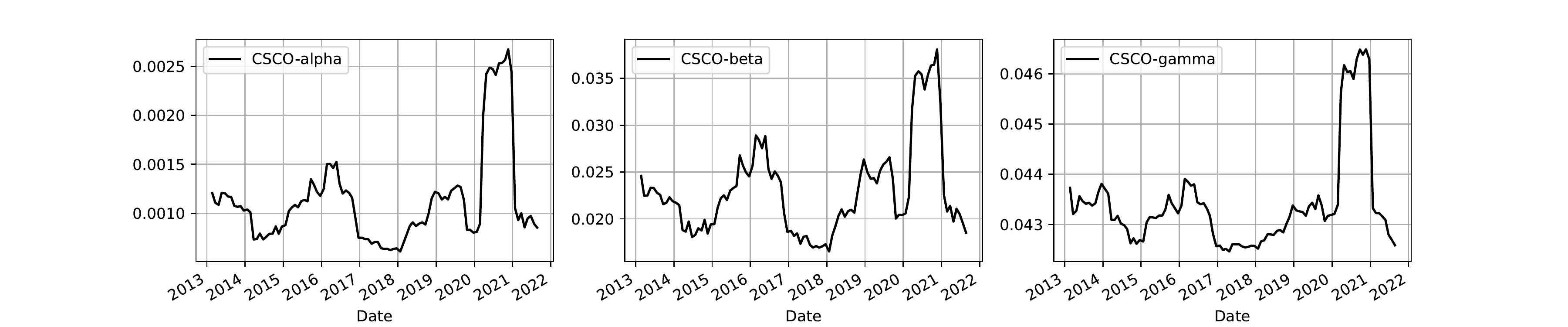}}
		\end{subfigure}
		\begin{subfigure}{1.0\textwidth}          \centerline{\includegraphics[width=1\linewidth]{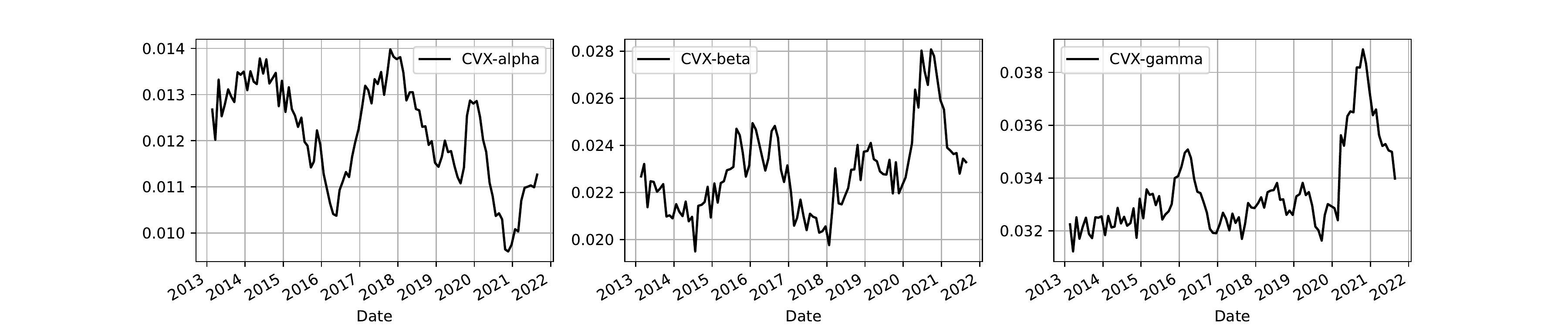}}
		\end{subfigure}
		\begin{subfigure}{1.0\textwidth}          \centerline{\includegraphics[width=1\linewidth]{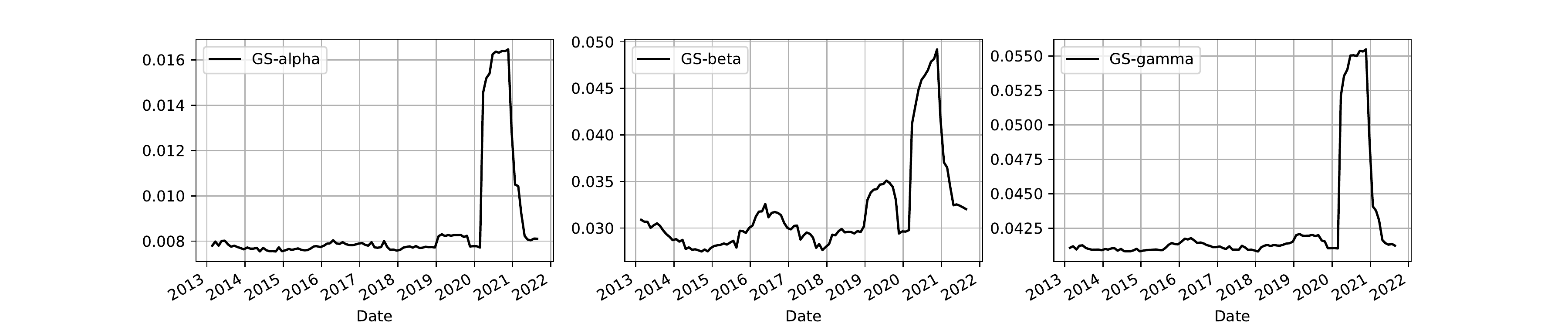}}
		\end{subfigure}
	\end{center}
	\caption{Predicted time-varying parameters of individual stock returns: BA, CAT, CSCO, CVX, and GS. The three curves in each row present the dynamics of $\alpha_i^t$ (left), $\beta_i^t$ (middle), and $\gamma_i^t$ (right), predicted by our GF-AGRU model.}
	\label{fig:param-fig-5stocks-2}
\end{figure}

To have an intuitive view on the temporal behaviors of time-varying parameters $\Theta^t_{\text{GF}}$ for the market returns and individual stock returns predicted by the proposed GF-AGRU model, we plot the curves of the predicted parameters over time and show the results of S\&P 500 and 10 representative stocks, as displayed in Figure \ref{fig:param-fig-market}, \ref{fig:param-fig-5stocks-1}, and \ref{fig:param-fig-5stocks-2}.

The left subplot in Figure \ref{fig:param-fig-market} shows the forecasts of $\alpha_M^t$ for the market returns (S\&P 500 index), while the right subplot is the corresponding $\beta_M^t$ curve. 
The $\alpha_M^t$ curve falls to the bottom at the end of 2015 and the end of 2018, and becomes volatile in the year of 2019 and 2020. Notice that the $\beta_M^t$ curve essentially mimics the volatility dynamics of market returns, and becomes significantly more volatile from the end of 2018.
We also give the predicted parameter curves for the individual stock returns in Figure \ref{fig:param-fig-5stocks-1} and \ref{fig:param-fig-5stocks-2}. The three curves in each row present the dynamics of $\alpha_i^t$ (left), $\beta_i^t$ (middle), and $\gamma_i^t$ (right), predicted by our GF-AGRU model. From the results of AAPL, we see that its $\alpha_i^t$ is much larger than those of other stocks. This is consistent with the intuition that AAPL is one of the best-performing stocks. For many stocks, the $\beta_i^t$ curve suffers from a spike in the year of 2020, consistent with the market crash caused by the pandemic. Similar patterns are observed in the $\gamma_i^t$ curves.

\section{Coverage Tests}
\label{AppendixB}
\addcontentsline{toc}{section}{Appendices}
\renewcommand{\thesubsection}{\Alph{subsection}}

\begin{table}
	\centering
	\caption{The three coverage tests for 99\%-VaR estimations of each stock given by GF-AGRU model.}
	\label{tab:coverage-test}
	\footnotesize
	\begin{tabular}{ccccc}
		\hline
		Stock & \makecell[c]{Number of\\ Violations} & \makecell[c]{$p$-value\\ of POF} &  \makecell[c]{$p$-value\\ of CCI} & \makecell[c]{$p$-value\\ of CC}  \\
		\hline
		S\&P 500 &        1 &    0.9762 &    0.8881 &   0.9897 \\
		AXP &        2 &     0.395 &    0.7773 &   0.6691 \\
		XOM &        3 &    0.1129 &     0.055 &   0.0452 \\
		AAPL &        0 &    0.1502 &       1.0 &   0.3552 \\
		BA &        1 &    0.9762 &    0.8881 &   0.9897 \\
		CAT &        0 &    0.1502 &       1.0 &   0.3552 \\
		CSCO &        0 &    0.1502 &       1.0 &   0.3552 \\
		CVX &        3 &    0.1129 &    0.6698 &     0.26 \\
		GS &        1 &    0.9762 &    0.8881 &   0.9897 \\
		HD &        2 &     0.395 &    0.7773 &   0.6691 \\
		PFE &        1 &    0.9762 &    0.8881 &   0.9897 \\
		IBM &        2 &     0.395 &    0.7773 &   0.6691 \\
		INTC &        0 &    0.1502 &       1.0 &   0.3552 \\
		JNJ &        2 &     0.395 &    0.7773 &   0.6691 \\
		KO &        1 &    0.9762 &    0.8881 &   0.9897 \\
		JPM &        1 &    0.9762 &    0.8881 &   0.9897 \\
		MCD &        1 &    0.9762 &    0.8881 &   0.9897 \\
		MMM &        1 &    0.9762 &    0.8881 &   0.9897 \\
		MRK &        0 &    0.1502 &       1.0 &   0.3552 \\
		MSFT &        0 &    0.1502 &       1.0 &   0.3552 \\
		NKE &        0 &    0.1502 &       1.0 &   0.3552 \\
		PG &        1 &    0.9762 &    0.8881 &   0.9897 \\
		TRV &        1 &    0.9762 &    0.8881 &   0.9897 \\
		UNH &        0 &    0.1502 &       1.0 &   0.3552 \\
		RTX &        1 &    0.9762 &    0.8881 &   0.9897 \\
		VZ &        0 &    0.1502 &       1.0 &   0.3552 \\
		WBA &        0 &    0.1502 &       1.0 &   0.3552 \\
		WMT &        1 &    0.9762 &    0.8881 &   0.9897 \\
		DIS &        2 &     0.395 &    0.7773 &   0.6691 \\
		\hline
		\makecell[c]{Rejection\\ Count ($<0.05$)} &          &         0 &         0 &        1 \\
		\hline
	\end{tabular}
\end{table}

\begin{table}
	\centering
	\caption{The three coverage tests for 95\%-VaR estimations of each stock given by GF-AGRU model.}
	\label{tab:coverage-test-95}
	\footnotesize
	\begin{tabular}{ccccc}
		\hline
		Stock & \makecell[c]{Number of\\ Violations} & \makecell[c]{$p$-value\\ of POF} &  \makecell[c]{$p$-value\\ of CCI} & \makecell[c]{$p$-value\\ of CC} \\
		\hline
		S\&P 500 &        6 &    0.7077 &    0.3303 &   0.5803 \\
		AXP &        5 &    0.9457 &    0.2114 &   0.4569 \\
		XOM &        7 &    0.4266 &    0.4729 &   0.5635 \\
		AAPL &        1 &    0.0226 &    0.8881 &   0.0737 \\
		BA &        5 &    0.9457 &    0.2114 &   0.4569 \\
		CAT &        4 &     0.589 &    0.5677 &    0.734 \\
		CSCO &        1 &    0.0226 &    0.8881 &   0.0737 \\
		CVX &        5 &    0.9457 &    0.4727 &   0.7709 \\
		GS &        3 &    0.2933 &     0.055 &   0.0913 \\
		HD &        2 &    0.1057 &    0.7773 &   0.2596 \\
		PFE &        4 &     0.589 &    0.1193 &   0.2569 \\
		IBM &        6 &    0.7077 &    0.3863 &   0.6405 \\
		INTC &        1 &    0.0226 &    0.8881 &   0.0737 \\
		JNJ &        5 &    0.9457 &    0.4727 &   0.7709 \\
		KO &        2 &    0.1057 &    0.7773 &   0.2596 \\
		JPM &        3 &    0.2933 &    0.6698 &   0.5256 \\
		MCD &        3 &    0.2933 &    0.6698 &   0.5256 \\
		MMM &        5 &    0.9457 &    0.2114 &   0.4569 \\
		MRK &        2 &    0.1057 &    0.7773 &   0.2596 \\
		MSFT &        1 &    0.0226 &    0.8881 &   0.0737 \\
		NKE &        2 &    0.1057 &    0.7773 &   0.2596 \\
		PG &        3 &    0.2933 &    0.6698 &   0.5256 \\
		TRV &        2 &    0.1057 &    0.7773 &   0.2596 \\
		UNH &        0 &    0.0012 &       1.0 &   0.0051 \\
		RTX &        6 &    0.7077 &    0.3303 &   0.5803 \\
		VZ &        2 &    0.1057 &    0.7773 &   0.2596 \\
		WBA &        6 &    0.7077 &    0.3303 &   0.5803 \\
		WMT &        2 &    0.1057 &    0.7773 &   0.2596 \\
		DIS &        2 &    0.1057 &    0.7773 &   0.2596 \\
		\hline
		\makecell[c]{Rejection\\ Count ($<0.05$)} &          &         5 &         0 &        1 \\
		\hline
	\end{tabular}
\end{table}

\begin{table}
	\centering
	\caption{The three coverage tests for 90\%-VaR estimations of each stock given by GF-AGRU model.}
	\label{tab:coverage-test-90}
	\footnotesize
	\begin{tabular}{ccccc}
		\hline
		Stock & \makecell[c]{Number of\\ Violations} & \makecell[c]{$p$-value\\ of POF} &  \makecell[c]{$p$-value\\ of CCI} & \makecell[c]{$p$-value\\ of CC}  \\
		\hline
		S\&P 500 &       11 &    0.8199 &    0.4352 &   0.7187 \\
		AXP &        6 &    0.1284 &    0.3303 &    0.196 \\
		XOM &       10 &    0.9212 &    0.9825 &   0.9949 \\
		AAPL &        7 &    0.2526 &    0.4729 &   0.4017 \\
		BA &       11 &    0.8199 &    0.8447 &   0.9559 \\
		CAT &        7 &    0.2526 &    0.3096 &   0.3102 \\
		CSCO &        4 &    0.0195 &    0.1193 &   0.0195 \\
		CVX &        6 &    0.1284 &    0.3303 &    0.196 \\
		GS &        7 &    0.2526 &    0.0628 &    0.092 \\
		HD &        3 &    0.0053 &    0.6698 &   0.0189 \\
		PFE &        5 &    0.0554 &     0.159 &   0.0592 \\
		IBM &       10 &    0.9212 &    0.9825 &   0.9949 \\
		INTC &        8 &    0.4335 &    0.6338 &   0.6569 \\
		JNJ &        8 &    0.4335 &     0.276 &   0.4066 \\
		KO &        5 &    0.0554 &    0.4727 &   0.1233 \\
		JPM &        6 &    0.1284 &    0.3863 &   0.2163 \\
		MCD &        5 &    0.0554 &    0.4727 &   0.1233 \\
		MMM &        9 &    0.6633 &    0.1517 &   0.3256 \\
		MRK &        3 &    0.0053 &    0.6698 &   0.0189 \\
		MSFT &        1 &    0.0001 &    0.8881 &   0.0006 \\
		NKE &        5 &    0.0554 &    0.2114 &   0.0731 \\
		PG &        5 &    0.0554 &    0.2114 &   0.0731 \\
		TRV &        4 &    0.0195 &    0.5677 &   0.0556 \\
		UNH &        4 &    0.0195 &    0.1193 &   0.0195 \\
		RTX &       10 &    0.9212 &    0.3021 &   0.5843 \\
		VZ &        3 &    0.0053 &    0.6698 &   0.0189 \\
		WBA &        9 &    0.6633 &    0.8062 &   0.8826 \\
		WMT &        7 &    0.2526 &    0.4729 &   0.4017 \\
		DIS &        5 &    0.0554 &    0.0113 &   0.0065 \\
		\hline
		\makecell[c]{Rejection\\ Count ($<0.05$)} &          &         7 &         1 &        7 \\
		\hline
	\end{tabular}
\end{table}

We test how well the proposed model predicts the return distribution of each stock, especially focusing on the tail risk estimations. The non-linear transformation in our generative factor model provides flexible heavy-tail properties for stock returns. To be more specific, we conduct some coverage tests to backtest the predictions of VaR at 99\%, 95\%, or 90\% level for each stock.

In detail, we utilize the generative mechanism of the proposed GF-AGRU model and simulate $n=10,000$ samples at each prediction date $t$ using the predicted parameters of the generative factor model. Then for each stock (as well as S\&P 500), we estimate VaR from these samples with confidence level $q=0.99,0.95,0.90$ and repeat the procedure over the out-of-sample set to form a sequence of VaR estimations with a total length $L=103$. 
We apply the classic POF unconditional coverage test \citep{kupiec1995techniques} to test the proportion of violations of VaR predictions (a violation means an observation is higher than the VaR prediction of the loss), which employs a likelihood ratio to assess the consistency between the violation probability and the probability implied by the VaR confidence level. Besides, we also check  whether the probability of observing a new violation depends on the past violation occurrence, i.e., conducting the independence test. For this purpose, we utilize the likelihood-ratio-based Conditional Coverage Independence (CCI) test \citep{christoffersen1998evaluating}. At last, we adopt a unified test combining the unconditional coverage test and the independence test, denoted as the Conditional Coverage (CC) test \citep{dias2013market}.

Table \ref{tab:coverage-test} shows the number of violations of our VaR estimations against the real monthly returns for each stock, together with the $p$-value of each test, under the confidence level $q=0.99$. In such a case, the expected number of violations shall be around 1 and we see that the maximum number of violations is 3. The three coverage tests also scarcely lead to rejections of the null hypothesis ($p$-value is smaller than 0.05) and we conclude that our GF-AGRU model provides robust 99\%-VaR estimations and performs well in tail risk coverage.
Table \ref{tab:coverage-test-95} and Table \ref{tab:coverage-test-90} are coverage test results of VaR estimations under the confidence level $q=0.95$ and $q=0.90$, respectively. They also show acceptable test results with small rejection counts.

\section{Robustness Analysis}
\label{AppendixC}
\addcontentsline{toc}{section}{Appendices}
\renewcommand{\thesubsection}{\Alph{subsection}}

We check the robustness of our model across various settings or under some inevitable randomness.
We mainly consider three cases of robustness analysis in line with the threads of our numerical experiment, covering the model hyper-parameter specification, the randomness in training the neural network, and the randomness of simulation in CVaR optimization. 
Accordingly, in the following, we address these robustness concerns separately.  

\begin{sidewaystable}
	\caption{Robustness of the GF-AGRU model in different hidden dimension specifications, by exhibiting the investment performance of CVaR portfolio optimization on the Dow Jones stocks with different settings of confidence level $q$ and sample size $n$. Three specifications are considered: the default hidden dimensions $D_h^M=4, D_h^i = 6$ (Origin), the Hidden-less setting $D_h^M=3, D_h^i = 5$, and the Hidden-more setting $D_h^M=5, D_h^i = 7$. We still include the EW method as a benchmark. $R_1$, $R_2$, and $R_3$ represent the target returns of $0.01$, $0.02$, and $0.03$, respectively. $R_{\text{ew}}$ represents the time-varying target return achieved by EW. Given a specific target return, the best performance under each evaluation metric is displayed with bold.} 
	\scriptsize
	\centering
	
	\begin{tabular}{cccccccccccccccc}
		\hline
		\multicolumn{16}{l}{\textbf{Panel A: CVaR Optimization with $q=0.90$, $n=5000$}} \\
		\hline
		Metric & EW & \multicolumn{4}{c}{Origin} && \multicolumn{4}{c}{Hidden-less} && \multicolumn{4}{c}{Hidden-more}    \\ 
		\cline{3-6} \cline{8-11} \cline{13-16}   
		& &  $R_1$ &$R_2$  &$R_3$    & $R_{\text{ew}} $   &&  
		$R_1$ &$R_2$  &$R_3$    & $R_{\text{ew}} $   && 
		$R_1$ &$R_2$  &$R_3$    & $R_{\text{ew}} $   \\
		\hline
		AV &  0.151 &  0.243 &           0.311 &  \textbf{0.318} &           0.334 &   &  \textbf{0.312} &           0.258 &             0.3 &  \textbf{0.358} &   &           0.226 &  \textbf{0.313} &  \textbf{0.318} &  0.281 \\
		SD &  0.139 &  0.171 &           0.183 &           0.238 &   \textbf{0.22} &   &  \textbf{0.167} &  \textbf{0.172} &  \textbf{0.237} &   \textbf{0.22} &   &           0.173 &           0.181 &           0.247 &  0.233 \\
		IR &  1.088 &   1.42 &           1.698 &  \textbf{1.337} &           1.517 &   &   \textbf{1.87} &             1.5 &           1.266 &  \textbf{1.627} &   &           1.311 &  \textbf{1.725} &           1.287 &  1.207 \\
		MD &  0.256 &  0.202 &  \textbf{0.196} &  \textbf{0.236} &           0.202 &   &   \textbf{0.17} &           0.201 &           0.245 &  \textbf{0.139} &   &           0.198 &           0.209 &           0.243 &   0.19 \\
		ES &  1.133 &    1.0 &  \textbf{1.053} &  \textbf{1.491} &           1.155 &   &  \textbf{0.915} &           1.117 &           1.557 &  \textbf{1.082} &   &           0.983 &           1.065 &           1.605 &  1.305 \\
		SK & -1.610 &  0.213 &   \textbf{0.21} &  \textbf{0.277} &  \textbf{1.314} &   &          -0.142 &          -0.346 &           0.167 &           1.305 &   &  \textbf{0.999} &           -0.01 &           0.157 &   1.12 \\
		CR &  0.133 &  0.243 &  \textbf{0.295} &  \textbf{0.213} &           0.289 &   &  \textbf{0.341} &           0.231 &           0.193 &  \textbf{0.331} &   &            0.23 &           0.294 &           0.198 &  0.215 \\
		RR &  0.191 &  0.307 &  \textbf{0.361} &  \textbf{0.275} &           0.355 &   &   \textbf{0.41} &           0.294 &           0.254 &  \textbf{0.399} &   &           0.293 &            0.36 &           0.259 &  0.277 \\

	\end{tabular}
	
	\begin{tabular}{cccccccccccccccc}
		\hline
		\multicolumn{16}{l}{\textbf{Panel B: CVaR Optimization with $q=0.95$, $n=10000$}} \\
		\hline
		Metric & EW & \multicolumn{4}{c}{Origin} && \multicolumn{4}{c}{Hidden-less} && \multicolumn{4}{c}{Hidden-more}    \\ 
		\cline{3-6} \cline{8-11} \cline{13-16}     
		& &  $R_1$ &$R_2$  &$R_3$    & $R_{\text{ew}} $   &&      
		$R_1$ &$R_2$  &$R_3$    & $R_{\text{ew}} $  && 
		$R_1$ &$R_2$  &$R_3$    & $R_{\text{ew}} $    \\
		\hline
		AV &  0.151 &           0.242 &  \textbf{0.318} &  \textbf{0.323} &  0.328 &   &  \textbf{0.319} &          0.254 &           0.299 &  \textbf{0.353} &   &           0.213 &           0.302 &  0.314 &  0.292 \\
		SD &  0.139 &           0.167 &           0.183 &           0.238 &  0.222 &   &           0.166 &  \textbf{0.17} &  \textbf{0.236} &  \textbf{0.219} &   &  \textbf{0.154} &           0.173 &  0.244 &  0.235 \\
		IR &  1.088 &           1.452 &           1.739 &  \textbf{1.357} &   1.48 &   &  \textbf{1.922} &          1.493 &           1.269 &  \textbf{1.608} &   &           1.385 &  \textbf{1.747} &   1.29 &  1.242 \\
		MD &  0.256 &           0.202 &  \textbf{0.196} &  \textbf{0.234} &  0.202 &   &   \textbf{0.17} &          0.201 &            0.25 &  \textbf{0.139} &   &            0.22 &           0.245 &  0.245 &   0.19 \\
		ES &  1.133 &             1.0 &  \textbf{1.053} &  \textbf{1.483} &  1.159 &   &  \textbf{0.915} &           1.11 &           1.542 &  \textbf{1.095} &   &           0.987 &           1.156 &  1.588 &  1.291 \\
		SK & -1.610 &  \textbf{0.235} &  \textbf{0.162} &  \textbf{0.245} &  1.276 &   &          -0.164 &         -0.356 &           0.079 &  \textbf{1.317} &   &          -0.416 &          -0.623 &  0.046 &  1.074 \\
		CR &  0.133 &           0.242 &  \textbf{0.302} &  \textbf{0.218} &  0.283 &   &  \textbf{0.348} &          0.229 &           0.194 &  \textbf{0.322} &   &           0.215 &           0.261 &  0.198 &  0.226 \\
		RR &  0.191 &           0.306 &  \textbf{0.368} &   \textbf{0.28} &  0.349 &   &  \textbf{0.417} &          0.292 &           0.255 &   \textbf{0.39} &   &           0.277 &           0.326 &  0.259 &  0.289 \\

	\end{tabular}
	
	\begin{tabular}{cccccccccccccccc}
		\hline
		\multicolumn{16}{l}{\textbf{Panel C: CVaR Optimization with $q=0.99$, $n=50000$}} \\
		\hline
		Metric & EW & \multicolumn{4}{c}{Origin} && \multicolumn{4}{c}{Hidden-less} && \multicolumn{4}{c}{Hidden-more}    \\ 
		\cline{3-6} \cline{8-11} \cline{13-16}     
		& &  $R_1$ &$R_2$  &$R_3$    & $R_{\text{ew}} $  &&      
		$R_1$ &$R_2$  &$R_3$    & $R_{\text{ew}} $  && 
		$R_1$ &$R_2$  &$R_3$    & $R_{\text{ew}} $  \\
		\hline
		AV &  0.151 &           0.218 &  \textbf{0.317} &  \textbf{0.324} &           0.309 &   &  \textbf{0.301} &   0.271 &  0.307 &  \textbf{0.339} &   &           0.172 &           0.265 &  0.302 &          0.286 \\
		SD &  0.139 &           0.159 &           0.183 &  \textbf{0.238} &   \textbf{0.22} &   &           0.171 &   0.179 &  0.239 &           0.224 &   &  \textbf{0.145} &   \textbf{0.16} &  0.242 &           0.23 \\
		IR &  1.088 &           1.372 &  \textbf{1.737} &   \textbf{1.36} &           1.402 &   &  \textbf{1.762} &   1.521 &  1.286 &  \textbf{1.514} &   &           1.182 &           1.659 &  1.247 &          1.244 \\
		MD &  0.256 &           0.202 &  \textbf{0.196} &  \textbf{0.234} &           0.202 &   &           0.198 &   0.202 &  0.252 &             0.2 &   &  \textbf{0.191} &           0.204 &  0.249 &  \textbf{0.19} \\
		ES &  1.133 &           0.971 &            1.05 &  \textbf{1.478} &  \textbf{1.158} &   &           1.007 &   1.115 &  1.561 &           1.214 &   &  \textbf{0.948} &  \textbf{1.049} &  1.565 &          1.262 \\
		SK & -1.610 &  \textbf{0.216} &  \textbf{0.173} &   \textbf{0.26} &  \textbf{1.359} &   &          -0.068 &  -0.171 &  0.102 &           1.103 &   &           0.006 &           -0.41 &  0.125 &          1.159 \\
		CR &  0.133 &           0.225 &  \textbf{0.302} &  \textbf{0.219} &           0.267 &   &  \textbf{0.299} &   0.243 &  0.196 &  \textbf{0.279} &   &           0.181 &           0.253 &  0.193 &          0.226 \\
		RR &  0.191 &           0.287 &  \textbf{0.369} &  \textbf{0.281} &           0.331 &   &  \textbf{0.366} &   0.307 &  0.258 &  \textbf{0.345} &   &           0.241 &           0.317 &  0.254 &          0.289 \\

		\hline
		
	\end{tabular}
	
	
	\label{hyparam-dow}
\end{sidewaystable}

\begin{table}
	\centering
	\caption{Robustness in training randomness of the GF-AGRU model, by exhibiting the investment performance of CVaR portfolio optimization on the Dow Jones stocks with different settings of confidence level $q$ and sample size $n$. 
		$R_1$, $R_2$, and $R_3$ represent the target returns of $0.01$, $0.02$, and $0.03$, respectively. $R_{\text{ew}}$ represents the time-varying target return achieved by EW. The results reported are the averages from 6 groups of trainings and the corresponding standard deviations are shown in brackets.}
	\footnotesize
	\begin{tabular}{cccccc}
		\hline
		\multicolumn{5}{l}{\textbf{Panel A: CVaR Optimization with} $q=0.90, n=5000$} \\
		\hline
		& $R_1$ &$R_2$  &$R_3$    & $R_{\text{ew}}$\\
		\hline
		AV &  0.261 (0.0315) &  0.296 (0.0243) &  0.313 (0.0083) &  0.329 (0.0128) \\
		SD &  0.176 (0.0049) &  0.179 (0.0069) &    0.239 (0.01) &  0.224 (0.0035) \\
		IR &  1.481 (0.1606) &  1.657 (0.1289) &  1.314 (0.0685) &  1.464 (0.0605) \\
		MD &   0.199 (0.004) &    0.2 (0.0028) &  0.242 (0.0169) &  0.204 (0.0105) \\
		ES &  1.021 (0.0328) &   1.06 (0.0295) &  1.536 (0.0755) &  1.211 (0.0553) \\
		SK &  0.212 (0.1947) &   0.026 (0.288) &  0.216 (0.2131) &  1.192 (0.1184) \\
		CR &  0.256 (0.0355) &    0.28 (0.031) &  0.204 (0.0134) &  0.272 (0.0207) \\
		RR &   0.32 (0.0373) &  0.345 (0.0325) &   0.266 (0.014) &  0.337 (0.0217) \\
	\end{tabular}
	
	
	\begin{tabular}{cccccc}
		\hline
		\multicolumn{5}{l}{\textbf{Panel B: CVaR Optimization with} $q=0.95, n=10000$} \\
		\hline
		& $R_1$ &$R_2$  &$R_3$    & $R_{\text{ew}}$\\
		\hline
		AV &   0.25 (0.0289) &  0.294 (0.0276) &  0.312 (0.0104) &  0.325 (0.0107) \\
		SD &  0.174 (0.0063) &  0.177 (0.0067) &  0.238 (0.0091) &  0.226 (0.0032) \\
		IR &  1.438 (0.1548) &  1.656 (0.1329) &  1.311 (0.0606) &  1.441 (0.0533) \\
		MD &  0.199 (0.0034) &    0.2 (0.0028) &  0.246 (0.0163) &   0.198 (0.005) \\
		ES &  1.029 (0.0473) &  1.055 (0.0309) &  1.523 (0.0746) &   1.219 (0.057) \\
		SK &  0.197 (0.2229) &  0.008 (0.2643) &    0.181 (0.15) &     1.175 (0.1) \\
		CR &  0.244 (0.0369) &  0.279 (0.0346) &  0.206 (0.0151) &  0.267 (0.0164) \\
		RR &  0.307 (0.0388) &  0.345 (0.0364) &  0.267 (0.0159) &  0.332 (0.0173) \\
	\end{tabular}
	

	\begin{tabular}{cccccc}
		\hline
		\multicolumn{5}{l}{\textbf{Panel C: CVaR Optimization with} $q=0.99, n=50000$} \\
		\hline
		& $R_1$ &$R_2$  &$R_3$    & $R_{\text{ew}}$ \\
		\hline
		AV &  0.233 (0.0269) &  0.291 (0.0316) &  0.314 (0.0111) &  0.318 (0.0156) \\
		SD &  0.172 (0.0094) &  0.178 (0.0064) &   0.239 (0.009) &  0.228 (0.0065) \\
		IR &  1.357 (0.1482) &    1.63 (0.149) &  1.314 (0.0613) &  1.395 (0.0898) \\
		MD &  0.205 (0.0133) &  0.201 (0.0028) &  0.248 (0.0164) &  0.221 (0.0359) \\
		ES &  1.037 (0.0758) &   1.061 (0.026) &  1.538 (0.0746) &  1.286 (0.1247) \\
		SK &  0.204 (0.2372) &  0.004 (0.2433) &  0.185 (0.1399) &  1.029 (0.3159) \\
		CR &  0.226 (0.0358) &  0.275 (0.0359) &  0.204 (0.0147) &    0.25 (0.031) \\
		RR &  0.289 (0.0376) &   0.34 (0.0377) &  0.266 (0.0155) &  0.313 (0.0326) \\
		\hline
	\end{tabular}
	\label{train-robust-GF-AGRU}
\end{table}

\begin{table}
	\centering
	\caption{Robustness in simulation randomness for the GF-AGRU model, by exhibiting the investment performance of CVaR portfolio optimization on the Dow Jones stocks with different settings of confidence level $q$ and sample size $n$. 
		$R_1$, $R_2$, and $R_3$ represent the target returns of $0.01$, $0.02$, and $0.03$, respectively. $R_{\text{ew}}$ represents the time-varying target return achieved by EW. The results reported are the averages from 10 simulations \& optimizations and the corresponding standard deviations are shown in brackets.}
	\footnotesize
	\begin{tabular}{cccccc}
		\hline
		\multicolumn{5}{l}{\textbf{Panel A: CVaR Optimization with} $q=0.90, n=5000$} \\
		\hline
		& $R_1$ &$R_2$  &$R_3$    & $R_{\text{ew}}$ \\
		\hline
		AV &  0.243 (0.0121) &  0.311 (0.0038) &  0.318 (0.0026) &  0.334 (0.0041) \\
		SD &  0.171 (0.0013) &   0.183 (0.001) &  0.238 (0.0016) &   0.22 (0.0004) \\
		IR &   1.42 (0.0688) &  1.698 (0.0166) &  1.337 (0.0114) &  1.517 (0.0189) \\
		MD &     0.202 (0.0) &  0.196 (0.0006) &  0.236 (0.0032) &     0.202 (0.0) \\
		ES &       1.0 (0.0) &  1.053 (0.0054) &  1.491 (0.0078) &  1.155 (0.0072) \\
		SK &  0.213 (0.0364) &   0.21 (0.0413) &  0.277 (0.0607) &  1.314 (0.0175) \\
		CR &  0.243 (0.0121) &  0.295 (0.0044) &  0.213 (0.0017) &  0.289 (0.0039) \\
		RR &  0.307 (0.0127) &  0.361 (0.0046) &  0.275 (0.0018) &  0.355 (0.0041) \\
	\end{tabular}
	
	
	\begin{tabular}{cccccc}
		\hline
		\multicolumn{5}{l}{\textbf{Panel B: CVaR Optimization with} $q=0.95, n=10000$} \\
		\hline
		& $R_1$ &$R_2$  &$R_3$    & $R_{\text{ew}}$ \\
		\hline
		AV &  0.242 (0.0053) &  0.318 (0.0013) &  0.323 (0.0026) &  0.328 (0.0013) \\
		SD &   0.167 (0.001) &  0.183 (0.0004) &  0.238 (0.0007) &  0.222 (0.0002) \\
		IR &  1.452 (0.0245) &  1.739 (0.0065) &  1.357 (0.0106) &   1.48 (0.0056) \\
		MD &     0.202 (0.0) &  0.196 (0.0003) &  0.234 (0.0036) &     0.202 (0.0) \\
		ES &       1.0 (0.0) &  1.053 (0.0049) &  1.483 (0.0112) &  1.159 (0.0052) \\
		SK &  0.235 (0.0165) &  0.162 (0.0175) &  0.245 (0.0279) &  1.276 (0.0071) \\
		CR &  0.242 (0.0053) &  0.302 (0.0019) &  0.218 (0.0029) &  0.283 (0.0019) \\
		RR &  0.306 (0.0056) &   0.368 (0.002) &   0.28 (0.0031) &   0.349 (0.002) \\
	\end{tabular}
	

	\begin{tabular}{cccccc}
		\hline
		\multicolumn{5}{l}{\textbf{Panel C: CVaR Optimization with} $q=0.99, n=50000$} \\
		\hline
		& $R_1$ &$R_2$  &$R_3$    & $R_{\text{ew}}$ \\
		\hline
		AV &  0.218 (0.0006) &  0.317 (0.0004) &  0.324 (0.0006) &  0.309 (0.0021) \\
		SD &  0.159 (0.0001) &  0.183 (0.0002) &  0.238 (0.0002) &   0.22 (0.0001) \\
		IR &  1.372 (0.0035) &  1.737 (0.0028) &   1.36 (0.0029) &    1.402 (0.01) \\
		MD &     0.202 (0.0) &  0.196 (0.0001) &  0.234 (0.0019) &     0.202 (0.0) \\
		ES &     0.971 (0.0) &   1.05 (0.0017) &  1.478 (0.0049) &  1.158 (0.0021) \\
		SK &  0.216 (0.0027) &  0.173 (0.0062) &    0.26 (0.007) &  1.359 (0.0066) \\
		CR &  0.225 (0.0006) &  0.302 (0.0005) &   0.219 (0.001) &  0.267 (0.0017) \\
		RR &  0.287 (0.0006) &  0.369 (0.0006) &  0.281 (0.0011) &  0.331 (0.0018) \\
		\hline
	\end{tabular}
	\label{simu-robustness-GF-AGRU}
\end{table}

\subsection{Robustness in Hyper-parameter Specifications}

The flexibility in constructing a neural network-based model gives rise to a concern regarding the robustness in different hyper-parameter specifications. Regarding our GF-AGRU model, we focus on the primary hyper-parameters that impact the model performance most, that is, the hidden dimensions in GRU.
The default hidden dimension we have used for the market return fitting step is $D_{h}^M = 4$, whereas for individual stock fitting, the default hidden dimension is set as $D_{h}^i = 6$. These values are determined by us to be twice the output dimension of GRU, which is 2 in the market fitting step and 3 in the individual stock fitting. In detail, in Algorithm \ref{algorithm_N}, we anticipate obtaining time-varying forecasts of the model parameters $\Theta_M^t=\{\alpha_M^t,\beta_M^t\}$ as the outputs of TV-AGRU, which have a dimension 2. Similarly, in Algorithm \ref{algorithm_i} for individual stocks, the forecasting parameters $\Theta_i^t=\{\alpha_i^t,\beta_i^t,\gamma_i^t\}$ have a dimension 3. We thus set moderate dimensions for $D_{h}^M$ and $D_{h}^i$, to reduce the risk of overfitting or underfitting.

To further examine the impact of $D_{h}^M$ and $D_{h}^i$ (as the primary hyper-parameters), we introduce some perturbations to the two hidden dimensions and examine the portfolio performance. Based on some previous experience, we refrain from considering large hidden dimensions because both the input dimensions (2 or 4) and the output dimensions (2 or 3) of GRU are small. So, the two alternative specifications we will explore are: $D_h^M=3,D_h^i=5$ (referred to as the Hidden-less setting) and $D_h^M=5,D_h^i=7$ (referred to as the Hidden-more setting).
The investment performance displayed in Table \ref{hyparam-dow} demonstrates that compared with the default Origin setting, smaller hidden dimensions (Hidden-less) yield notable improvements under $R_1$ and $R_{\text{ew}}$, but slightly reduce the performance under $R_2$ and $R_3$. On the other hand, larger hidden dimensions (Hidden-more) seem to result in some moderate inferiority in performance. This confirms our suggestion that small hidden dimensions are preferred. Overall, minor adjustments to the hidden dimensions will cause some variations in the performance, but with no significant deterioration. Most importantly, the settings in Table \ref{hyparam-dow} all produce better performance than the benchmark models in Table \ref{benchmark2}.



\subsection{Robustness in Training Randomness}
The neural network model training procedure inherently introduces stochasticity, which may contribute to some variations in the resulting outcomes. Note that in our experiment, we do not employ random selection for batch splitting as many deep learning models do. Furthermore, in the gradient descent process for parameter optimization, we adopt the RMSProp optimizer provided by PyTorch, which is a deterministic optimizer that does not introduce additional randomness during training. Consequently, the only source of variability is attributed to the initialization of the neural network. 

Also, notice that we have utilized an ensemble averaging technique to reduce some randomness, but stochasticity still exists. Therefore, to investigate the robustness of the model's performance in training randomness, we examine  by repeating the training procedure multiple times and computing both the average and the variability of the performance. In detail, we select a total of 6 groups of random seeds, with each group consisting of $B_r=5$ different seeds (e.g., $\{10, 20, 30, 40, 50\}$). Each group is used to train the GF-AGRU model and then obtain the performance one time, as we do in the previous subsection. Then this will result in 6 distinct outcomes, for studying the robustness. We compute the mean  and the standard deviation of these six outcomes, as presented in Table \ref{train-robust-GF-AGRU}. 
We observe that the mean performance of the GF-AGRU model maintains its superiority over the benchmark models as well as the two variants, indicating consistent advantages. Regarding the variability, for most metrics, we see that the standard deviations in the brackets are smaller than $0.1$ in most considered scenarios, indicating adequate robustness in model training randomness. 

\subsection{Robustness in Simulation Randomness}

The GF-AGRU model is constructed with a generative approach, enabling it to simulate samples from the learned model. At each investment date within the test period, we simulate $n$ samples for the multivariate stock returns based on the Attention-GRU forecasts (averaged from $B_r=5$ independent trainings), where $n$ is the sample size corresponding to the confidence level $q$. Subsequently, we conduct mean-CVaR optimization with these samples. Given the inherent randomness associated with the generation of these samples, it is necessary to assess the robustness in simulation randomness. 

We perform the simulation of $n$ samples 10 times, and solve the CVaR optimization 10 times correspondingly, to report the average of the results. Actually, this is what we have exactly done in all above tables and figures.
Meanwhile, it is also essential to showcase the variation of the results observed across different simulations, and hence we provide the standard deviations in brackets in Table \ref{simu-robustness-GF-AGRU}. 
It is obviously observed that the standard deviations of these metrics are below $0.1$, and most of them are even smaller than $0.01$. The results indicate significantly reliable robustness in the simulation randomness in CVaR optimization.

\bibliographystyle{elsarticle-harv}

\bibliography{Bibliography-MM-MC}

\end{document}